\documentclass[aps,10pt,prd,twocolumn,groupedaddress,floatfix,longbibliography,superscriptaddress,notitlepage]{revtex4-2}

\usepackage{amsmath}
\usepackage{amssymb}
\usepackage{amsfonts}
\usepackage{bm}
\usepackage{times,float}
\usepackage{graphicx}
\usepackage[dvipsnames,svgnames]{xcolor}
\usepackage{hyperref}
\hypersetup{colorlinks=true, linkcolor=NavyBlue, citecolor=PineGreen,urlcolor=cyan}
\usepackage{physics}
\graphicspath{{Figures/}}
\usepackage{mathrsfs}
\usepackage{subcaption}

\newcommand{\partd}[3][1]{%
	\ifnum#1=1
	\frac{\partial #2}{\partial #3}%
	\else
	\frac{\partial^{#1}#2}{\partial #3^{#1}}%
	\fi
}

\newcommand{\ii}{\mathrm{i}}

\begin{document}
	
	\title{Non-Hermitian impurity scattering in graphene: Boltzmann transport and thermoelectric response}
	\author{Juan A. Ca\~nas}
	\email{juan.canas@correo.nucleares.unam.mx}
	\affiliation{Instituto de Ciencias Nucleares, Universidad Nacional Aut\'{o}noma de M\'{e}xico, 04510 Ciudad de M\'{e}xico, M\'{e}xico}
	\author{Daniel A. Bonilla}
	\email{daniel.bonillam@correo.nucleares.unam.mx}
	\affiliation{Instituto de Ciencias Nucleares, Universidad Nacional Aut\'{o}noma de M\'{e}xico, 04510 Ciudad de M\'{e}xico, M\'{e}xico}
	\author{A. Mart\'{i}n-Ruiz}
	\email{alberto.martin@nucleares.unam.mx}
	\affiliation{Instituto de Ciencias Nucleares, Universidad Nacional Aut\'{o}noma de M\'{e}xico, 04510 Ciudad de M\'{e}xico, M\'{e}xico}
	
\begin{abstract}
We investigate charge and thermoelectric transport in monolayer graphene containing a dilute distribution of finite-range non-Hermitian scattering centers. The impurities are modeled as circular complex potentials, whose imaginary component describes local carrier loss or gain. By solving the Dirac scattering problem exactly within a partial-wave approach, we obtain the nonunitary scattering matrix and derive the transport and absorption cross sections, which separately characterize momentum relaxation and net carrier exchange with the environment. To connect the microscopic scattering problem with stationary transport, we formulate a semiclassical Boltzmann description in which an external reservoir compensates the equilibrium particle loss or gain. This leads to an effective relaxation time governed by both elastic momentum scattering and non-Hermitian flux exchange. Using the full energy-dependent relaxation time, we evaluate the Onsager coefficients and the resulting electrical conductivity, electronic thermal conductivity, Seebeck coefficient, Lorenz ratio, and electronic thermoelectric figure of merit. We find that weak gain increases the effective carrier lifetime and enhances both charge and heat conductivities, while absorption produces the opposite behavior. More importantly, gain enhances the magnitude of the thermopower and the electronic figure of merit, whereas loss suppresses them. The Lorenz ratio remains close to the Sommerfeld value, with non-Hermiticity mainly modifying its finite-temperature corrections. Our results show that non-Hermitian scattering provides an additional mechanism for controlling the energy dependence of carrier relaxation and, consequently, the thermoelectric response of graphene.
\end{abstract}
	
	\maketitle
	
\section{Introduction}

Graphene is a two-dimensional platform for charge and heat transport with tunable carrier density, disorder, and geometry. Its low-energy carriers obey a Dirac equation, and sublattice chirality constrains scattering from potentials and defects \cite{CastroNetoRMP2009,PeresRMP2010}. Electrostatic gating accesses electron- and hole-dominated regimes \cite{DasSarmaRMP2011}. Electrical and electronic thermal conductivities, thermopower, and their combination in the electronic thermoelectric figure of merit characterize charge, heat, and heat-to-electricity conversion when phonon transport is neglected.

Early measurements established the gate-voltage dependence and sign reversal of thermopower across charge neutrality \cite{ZuevPRL2009,WeiPRL2009}. Charged and resonant impurities modify it through screening, impurity states, and energy-dependent scattering \cite{Lofwander2007,Hwang_PRB2009,Inglot2015}. This sensitivity, expressed by the Mott relation in the degenerate regime, also underlies graphene's photothermoelectric response \cite{Massicotte2021}. Near charge neutrality, enhanced Lorenz ratios have been associated with a hydrodynamic Dirac fluid \cite{Crossno2016}, while bipolar diffusion with a finite gap has been proposed as an alternative \cite{Tu2023}. These regimes differ from impurity-controlled transport at finite doping, while graphene's large phonon thermal conductivity further limits its total thermoelectric figure of merit \cite{CC_Kim_2025}.

At finite carrier density, realistic adsorbates and strong short-range defects can form resonant states beyond the first Born approximation \cite{WehlingPRL2010,FerreiraPRB2011}. An inaccurate Born rate does not, however, invalidate a Boltzmann description \cite{KlosPRB2010}. For circular potentials, exact partial-wave solutions retain the dependence on angle, energy, and defect size \cite{NovikovPRB2007,WuPRB2014}; arrays of circular electrostatic potentials have also produced a graphene analogue of Mie scattering \cite{CaridadNatComm2016}. Exact scattering has been used in kinetic thermoelectric calculations, where pseudomagnetic resonances and finite-range scalar potentials show how defect geometry controls energy-dependent relaxation \cite{CBM_R2025,CBM_PB2026}. Near localization, disorder can instead enhance thermopower and power factor through physics beyond independent scattering \cite{Francis2026}.

A different situation arises when a defect exchanges particles with its environment. Eliminating environmental degrees of freedom can generate a non-Hermitian effective operator, while a complex local potential phenomenologically describes absorption or amplification \cite{Muga2004,Ashida2020}. Non-Hermitian operators also arise in quasiparticle descriptions with unequal decay rates and exceptional points \cite{kozii_Arx2017}. Here, the imaginary potential makes the scattering matrix nonunitary because incoming and outgoing carrier fluxes need not be equal.

Localized non-Hermitian defects can modify spectra and propagation. Lossy or complex impurities redistribute the local density of states and alter bound-state formation \cite{Sukhachov2020,Kokkinakis2026}; non-Hermitian Dirac interfaces can show anomalous transmission \cite{Terh2023}; and imaginary disorder can support localized eigenstates together with long-range wave-packet spreading \cite{Tzortzakakis2021,Li2025}. Stationary transport additionally requires a prescription for particle supply and removal. In reservoir-based approaches, absorption must be accompanied by reinjection when particle conservation is imposed \cite{Brouwer1997}; local loss can selectively affect transport resonances and nonlinear current--voltage characteristics \cite{Visuri2022,Visuri2023}; lossy wires can be related to multiterminal transport \cite{Uchino2022}; and interactions introduce further dependence on loss and replenishment \cite{Gievers2024}.

Recent response theories make this dependence explicit. Generalized Kubo and Green-function formulations include normalization changes or source terms from nonunitary dynamics \cite{Sticlet2022,Yan2024}. Multiprobe schemes address gauge invariance with additional currents \cite{Wei2025}, while Lindblad--Keldysh extensions of Landauer--B\"uttiker transport include particle and energy exchange with gain and loss reservoirs \cite{Yang2026}. The interpretation of the effective non-Hermitian Hamiltonian also constrains the Green functions and occupations used for response calculations \cite{Kleger2026}. Stationary transport therefore depends on both the complex potential and the assumed occupations and particle-exchange prescription.

Thermoelectric effects of gain and loss depend on geometry and system. In superconducting double quantum dots, balanced complex potentials can reverse the Seebeck coefficient \cite{Niu2026}, while a finite complex barrier in graphene can enhance the electronic figure of merit under absorptive loss \cite{Bonilla_PRB2026}. A dilute impurity ensemble raises a distinct question: how does particle exchange at individual scatterers modify the energy-dependent relaxation of a macroscopic current? Single-barrier transmission alone does not determine this relaxation for an impurity ensemble.

Here we study electron-doped monolayer graphene with randomly distributed circular complex potentials. Motivated by recent developments in non-Hermitian transport, where gain and loss can qualitatively modify scattering, spectral properties, and stationary currents, we address how such effects can be incorporated into disorder-limited diffusive transport. In particular, we propose a non-Hermitian extension of the Boltzmann description for dilute, independent impurities by combining the exact single-defect partial-wave solution with a phenomenological kinetic equation. The scattering problem separates momentum redistribution from net flux exchange, while the kinetic model specifies how both processes relax deviations from a reference distribution maintained by an external reservoir. This provides an effective energy-dependent relaxation rate from which we calculate the electrical and electronic thermal conductivities, thermopower, and electronic figure of merit, and compare gain and loss with the Hermitian limit. The approach therefore extends the standard impurity-scattering framework to situations in which individual defects can exchange particles with their environment, providing a simple route for exploring non-Hermitian effects in semiclassical transport. The model is necessarily phenomenological: the complex potential alone does not determine reservoir occupations, the microscopic mechanism of particle exchange, or the conditions required for a stable stationary state. A fully microscopic open-system treatment would be needed to determine these ingredients self-consistently. The advantage of the present formulation is its direct applicability as an effective model, since non-Hermitian scattering can be incorporated into a Boltzmann transport calculation without introducing an explicit microscopic reservoir. Our predictions are consequently conditional on the assumed particle-exchange prescription and are restricted to diffusive transport away from charge neutrality and to the electronic subsystem, excluding phonon heat transport and the energetic cost of maintaining the reservoir.

This paper is organized as follows. Section~\ref{sec:Model} introduces the model, Sec.~\ref{sec:NHScattering} develops the scattering solution, and Sec.~\ref{sec:Transport} formulates the kinetic description and thermoelectric coefficients. Results and conclusions are presented in Secs.~\ref{sec:results_discussion} and \ref{sec:conclusions}.

\section{The Model: Non-Hermitian impurities}
\label{sec:Model}
	
	At low energies, charge carriers in monolayer graphene behave as chiral, massless Dirac fermions. Neglecting intervalley scattering, the continuum effective Hamiltonian around a single valley node $\chi = \pm 1$ is given by~\cite{CastroNetoRMP2009}
	\begin{equation}
		\hat{\mathcal{H}}_0 = \chi v_F \hat{\bm{\sigma}} \cdot \bm{p},
		\label{eq:H0_Dirac}
	\end{equation}
	where $v_F \approx 9.8 \times 10^{5}$~m/s is the Fermi velocity, $\bm{p} = -i\hbar\bm{\nabla}$ is the momentum operator measured relative to the Dirac point, and $\hat{\bm{\sigma}} = (\sigma_x, \sigma_y)$ is the vector of Pauli matrices acting on the sublattice (pseudospin) degrees of freedom. The dispersion relation is symmetric around the charge neutrality point, yielding the linear spectrum $\mathscr{E}_{s\chi}(\bm{q}) = s \chi \hbar v_F q$, where $s \chi = \pm 1$ denotes the conduction and valence bands, respectively~\cite{CastroNetoRMP2009}. In the following, we focus on electron-doped systems in the conduction band ($s\chi=+1$) with energy $E = \hbar v_F k > 0$. The total charge and energy currents will incorporate a factor of $g_s g_v = 4$ to account for the standard spin and valley degeneracies.
	
	To describe a realistic device with smooth disorder, such as that induced by screened charged impurities~\cite{CBM_PB2026} or local strain fields~\cite{CBM_R2025}, isolated defects are modeled as finite-range scalar potentials rather than zero-range delta functions~\cite{HwangPRL2007,NovikovPRB2007}. In this work, we extend this phenomenological framework to open systems where charge conservation is locally broken via particle injection or extraction. We consider a dilute, random distribution of non-Hermitian circular disk potentials described by the complex scalar profile:
	\begin{equation}
		V(\bm{r}) = V_0 \Theta(a - r) = (U + iW) \Theta(a - r),
		\label{eq:NH_potential}
	\end{equation}
	where $\Theta(x)$ is the Heaviside step function, while $U, W \in \mathbb{R}$ represent the Hermitian and non-Hermitian potential strengths, respectively. 
	
	Motivated by the optical model of nuclear physics~\cite{Feshbach1954,Mott,CantoHussein2013}, where complex potentials are routinely employed to phenomenologically describe the absorption or emission of particles during scattering events, we introduce the imaginary component $W$ to capture local charge non-conservation. The addition of this term breaks the hermiticity of the full Hamiltonian, $\hat{\mathcal{H}} = \hat{\mathcal{H}}_0 + V(\bm{r})\sigma_0$. Formally, $W < 0$ corresponds to a local sink (particle absorption/loss), whereas $W > 0$ characterizes a local source (particle amplification/gain) within the defect region. This sign convention ($U + iW$) is strictly chosen so that stationary states evolving with the standard time-phase factor $e^{-iEt/\hbar}$ naturally decay under absorption when $\operatorname{Im}E < 0$. Notably, this is opposite to the standard nuclear optical model ($U - iW$) traditionally employed to describe absorptive reactions in three-dimensional scattering systems~\cite{Feshbach1954,Mott,CantoHussein2013}.
	
	The physical consequence of this non-Hermitian term is rigorously demonstrated through the continuity equation. For the stationary Dirac equation $[\chi v_F \hat{\bm{\sigma}} \cdot \bm{p} + V(\bm{r})\sigma_0]\Psi = E\Psi$, evaluating the time-dependent evolution $i\hbar\partial_t\Psi = \hat{\mathcal{H}}\Psi$ alongside its adjoint yields a kinetic contribution $-\bm{\nabla}\cdot\bm{j}$ and a non-vanishing potential contribution $2\operatorname{Im}(V_0)\rho/\hbar$. This defines the 2D Dirac analog of the modified continuity equation for Schrödinger particles with complex potentials~\cite{CantoHussein2013}:
	\begin{equation}
		\frac{\partial\rho}{\partial t} + \bm{\nabla}\cdot\bm{j} = \frac{2W}{\hbar} \Theta(a-r) \rho.
		\label{eq:continuity_disk}
	\end{equation}
	Integrating over a macroscopic disk of radius $R > a$ at steady state and applying Gauss's theorem produces the global flux-balance law:
	\begin{equation}
		\oint_{r=R} \bm{j}\cdot\hat{\bm{r}} \, dl = \frac{2W}{\hbar} \int_{r<a} \rho(\bm{r}) \, d^2r.
		\label{eq:flux_balance}
	\end{equation}
	This macroscopic flux balance explicitly confirms our convention: $W < 0$ dictates a net inward particle flux (sink), while $W > 0$ generates a net outward flux (source).
	
	Scattering from these centers fundamentally alters the propagation of quasiparticles. In the exterior region ($r > a$), the quasiparticles propagate freely with a real wave number $k = E / (\hbar v_F)$. Conversely, inside the non-Hermitian disk ($r < a$), the scalar potential remains constant, rendering the inner region exactly solvable. However, due to the breaking of hermiticity, the effective wave number inside the scatterer becomes explicitly complex:
	\begin{equation}
		q = k - \frac{U + iW}{\hbar v_F}.
		\label{eq:complex_q}
	\end{equation}
	To ensure the analytical tractability of the transport properties, we assume the ensemble of defects lies in the dilute limit, characterized by the condition $n_i a^2 \ll 1$, where $n_i$ represents the areal impurity density. This inequality ensures that the average separation between independent non-Hermitian centers is much larger than their spatial extent, suppressing multiple-scattering interference and validating a semi-classical description around the stationary reference distribution.

\section{Non-Hermitian Scattering Formalism}
\label{sec:NHScattering}

\subsection{Partial Wave Analysis}
\label{sec:PartialWaves}
	To solve the scattering problem exactly, we exploit the rotational symmetry of the circular non-Hermitian disk. In the continuum Dirac description, the total angular momentum operator along the $z$-axis commutes with the full Hamiltonian $\hat{\mathcal{H}}$ \cite{CastroNetoRMP2009,NovikovPRB2007,CBM_R2025,CBM_PB2026}. Consequently, the spinor wave functions can be decomposed into independent angular momentum channels labeled by the eigenvalue $\hbar m_j$, where $m_j = m + 1/2$ (with $m \in \mathbb{Z}$) represents the total angular momentum quantum number. For a given channel $m_j$, the stationary state ansatz in polar coordinates $(r, \theta)$ takes the form
	\begin{equation}
		\Psi_{m_j}(r, \theta) = \begin{pmatrix} 
			\phi_A(r) e^{i(m_j - 1/2)\theta} \\ 
			i \phi_B(r) e^{i(m_j + 1/2)\theta} 
		\end{pmatrix},
		\label{eq:Ansatz_PW}
	\end{equation}
	where $\phi_A(r)$ and $\phi_B(r)$ are the radial amplitudes corresponding to the $A$ and $B$ sublattices, respectively.
	
	In the exterior region ($r > a$), the total wave function can be expressed in terms of first and second kind bessel functions \cite{CBM_R2025,CBM_PB2026}. Here, however, is convenient to write it in the Hankel basis:
	\begin{align}
		\phi_A^{\text{out}}(r) &= C_{m_j} \left[H_{m_j - 1/2}^{(2)}(kr) + S_{m_j}(k) H_{m_j - 1/2}^{(1)}(kr)\right], \label{eq:phiA_out} \\
		\phi_B^{\text{out}}(r) &= C_{m_j} \left[H_{m_j + 1/2}^{(2)}(kr) + S_{m_j}(k) H_{m_j + 1/2}^{(1)}(kr)\right], \label{eq:phiB_out}
	\end{align}
	where $H_\nu^{(1)}(x)$ and $H_\nu^{(2)}(x)$ are the Hankel functions of the first and second kind respectively, $C_{m_j}$ fixes the incoming amplitude in the channel $m_j$, and $S_{m_j}(k)$ is the channel
	scattering matrix.
	
	Inside the defect area ($r < a$), the system is governed by the constant complex potential $V_0$. To ensure physical admissibility, the wave function must remain regular at the origin, which isolates the Bessel function of the first kind. The internal radial amplitudes are given by:
	\begin{align}
		\phi_A^{\text{in}}(r) &= D_{m_j} J_{m_j - 1/2}(qr), \label{eq:phiA_in} \\
		\phi_B^{\text{in}}(r) &= D_{m_j} J_{m_j + 1/2}(qr), \label{eq:phiB_in}
	\end{align}
	where $q$ is the complex wave number in eq.~\eqref{eq:complex_q}, and $D_{m_j}$ represents the internal transmission amplitude.
	
	The scattering coefficients are determined by imposing the continuity of the spinor wave function across the potential interface, $\Psi_{m_j}^{\text{in}}(a, \theta) = \Psi_{m_j}^{\text{out}}(a, \theta)$. This matching condition enables elmination of the constants $C_{m_j}$ and $D_{m_j}$, so we obtain the explicit analytical form of the scattering matrix elements:
	
	\begin{equation}
		S_{m_j} = -\frac{H_{m_j + 1/2}^{(2)}(ka) J_{m_j - 1/2}(qa) - H_{m_j - 1/2}^{(2)}(ka) J_{m_j + 1/2}(qa)}{H_{m_j + 1/2}^{(1)}(ka) J_{m_j - 1/2}(qa) - H_{m_j - 1/2}^{(1)}(ka) J_{m_j + 1/2}(qa)}.
		\label{eq:Smj_matrix}
	\end{equation}
	Rather than defining complex phase shifts, which suffer from branch-cut ambiguities and lose their intuitive geometric interpretation under non-Hermitian dynamics \cite{Moiseyev_B2011}, it is physically more transparent to analyze the scattering matrix elements $S_{m_j}$, defined above. 
	
	The non-Hermitian nature of the scalar potential introduces fundamental departures from conventional Dirac barriers. In the Hermitian limit, the wave number $q$ is strictly real, which guarantees the unitarity of the scattering matrix ($|S_{m_j}|^2 = 1$). In that regime, the scattering is purely elastic and the radial flux is conserved within each individual channel. Conversely, when $W \neq 0$, the complex nature of $q$ breaks the unitarity of the S-matrix ($|S_{m_j}|^2 \neq 1$) \cite{Ghaemi_PRA2021}, providing a microscopic connection to the global flux-balance law derived in Eq.~\eqref{eq:flux_balance}.
	
	The modulus $|S_{m_j}|^2$ serves as a precise indicator of local particle non-conservation within the $m_j$-th channel. In the absorptive regime ($W < 0$), the scattering matrix satisfies $|S_{m_j}|^2 < 1$. This indicates that the outgoing circular wave carries less flux than the incoming wave, tracking the local dissipation or capture of charge carriers by the defect~\cite{CantoHussein2013}. Within the amplifying regime ($W > 0$), the scattering matrix satisfies $|S_{m_j}|^2 > 1$. Here, the defect acts as a localized pump, injecting carriers into the graphene sheet and amplifying the partial outgoing wave.

\subsection{Non-Hermitian Cross-Sections}
\label{subsec:CrossSections}

	The macroscopic transport properties of the system are governed by the scattering observables, which can be analytically extracted from the asymptotic behavior of the wave functions. Far from the non-Hermitian defect ($kr \gg 1$), the outgoing cylindrical waves can be approximated using the asymptotic expansions of the Hankel functions. The scattered part of the spinor wave function assumes the standard two-dimensional asymptotic form \cite{NovikovPRB2007}:
	\begin{equation}
		\Psi_{\text{sc}}(r, \theta) \simeq \frac{e^{ikr}}{\sqrt{-ir}} 
		\begin{pmatrix} 
			1 \\ 
			e^{i\theta} 
		\end{pmatrix} 
		f(\theta),
		\label{eq:Psi_sc_asymptotic}
	\end{equation}
	where $f(\theta)$ is the angular scattering amplitude. By matching this limit with the partial-wave expansion in Eqs.~\eqref{eq:phiA_out} and \eqref{eq:phiB_out}, the scattering amplitude can be expressed directly in terms of the S-matrix elements derived in Eq.~\eqref{eq:Smj_matrix}~\cite{NovikovPRB2007}:
	\begin{equation}
		f(\theta) = \frac{1}{2 \ii \sqrt{\pi k}} \sum_{m_j = -\infty}^{\infty} (S_{m_j} - 1) e^{i(m_j - 1/2)\theta}.
		\label{eq:Scat_Amplitude}
	\end{equation}
	
	In a standard Hermitian system, the radial probability flux is globally conserved, leading to purely elastic scattering. However, the presence of the complex potential $W \neq 0$ acts as a local source or sink of quasiparticles. The total scattering cross-section $\sigma_{\text{sc}}$, which quantifies the total scattered probability flux, is obtained by integrating $|f(\theta)|^2$ over all angles. Due to the orthogonality of the angular modes, it reduces to a simple sum over the individual channels:
	\begin{equation}
		\sigma_{\text{sc}} = 2\int_0^{2\pi} |f(\theta)|^2 \, d\theta = \frac{1}{k} \sum_{m_j = -\infty}^{\infty} |S_{m_j} - 1|^2.
		\label{eq:sigma_sc}
	\end{equation}
	
	To accurately quantify the degree of particle non-conservation, we define the absorption (or reaction) cross-section $\sigma_{\text{abs}}$. By integrating the net radial flux flowing into (or out of) the defect region and applying the global flux-balance condition from Eq.~\eqref{eq:flux_balance}, the absorption cross-section is exactly expressed as the sum of the unitarity deficits of each partial wave~\cite{CantoHussein2013}:
	\begin{equation}
		\sigma_{\text{abs}} = \frac{1}{k} \sum_{m_j = -\infty}^{\infty} \left( 1 - |S_{m_j}|^2 \right).
		\label{eq:sigma_abs}
	\end{equation}
	This quantity has an immediate physical interpretation: it is strictly positive for a localized sink ($W < 0$, $|S_{m_j}|^2 < 1$), reflecting a net loss of carriers, and becomes negative for an active source ($W > 0$, $|S_{m_j}|^2 > 1$), signifying carrier injection into the graphene sheet. 
	
	The sum of the elastic and absorptive processes yields the total cross-section, $\sigma_{\text{tot}} = \sigma_{\text{sc}} + \sigma_{\text{abs}}$. Substituting Eqs.~\eqref{eq:sigma_sc} and \eqref{eq:sigma_abs}, we recover the two-dimensional Dirac analog of the generalized optical theorem~\cite{NovikovPRB2007,CantoHussein2013}:
	\begin{equation}
		\sigma_{\text{tot}} = \frac{2}{k} \sum_{m_j = -\infty}^{\infty} \left[ 1 - \operatorname{Re}(S_{m_j}) \right].
		\label{eq:sigma_tot}
	\end{equation}
	
	Equations~\eqref{eq:sigma_sc} through \eqref{eq:sigma_tot} establish the complete cross-sectional framework for single-impurity non-Hermitian scattering. By encapsulating both the elastic redirection of quasiparticles and the local breaking of charge conservation entirely within the complex S-matrix elements, this formalism bypasses the ambiguities of non-Hermitian phase shifts. With the microscopic scattering observables rigorously defined, we now proceed to evaluate how this localized non-unitarity dictates the macroscopic scattering rates and the corresponding electronic transport properties.

\section{Semiclassical Transport Analysis} 
\label{sec:Transport}
	
\subsection{Modified Boltzmann Equation and Effective Relaxation Time}
\label{subsec:EffectiveTime}
	
	To transition from single-impurity scattering to macroscopic transport, we must evaluate the relevant cross-sections that dictate the kinetic collision rates. In our non-Hermitian framework, the scattering dynamics naturally decouple into two distinct mechanisms: the elastic redistribution of momentum and the scalar loss or gain of carrier flux. The elastic redistribution is governed by the transport (or momentum-relaxation) cross-section $\sigma_{\text{tr}}$. Unlike the total elastic cross-section, $\sigma_{\text{tr}}$ weighs large-angle scattering more heavily by incorporating a geometric factor $(1 - \cos\theta)$, which accounts for the degradation of forward momentum along the transport direction. In the partial-wave basis, this geometric factor induces interference between adjacent angular momentum channels, yielding \cite{NovikovPRB2007}
    \begin{align}
        \sigma_{\text{tr}}(k) &= 2\int_0^{2\pi} |f(\theta)|^2 (1 - \cos\theta) , d\theta \notag\\[6pt]
        &= \frac{1}{2k} \sum_{m_j = -\infty}^{\infty} \left| S_{m_j}(k) - S_{m_j+1}(k) \right|^2.
        \label{eq:sigma_tr}
    \end{align}
    Conversely, the degree of non-unitarity is a purely scalar process independent of momentum redirection. This flux imbalance is captured entirely by the absorption cross-section $\sigma_{\text{abs}}(k)$ previously defined in Eq.~\eqref{eq:sigma_abs}. The corresponding macroscopic rates are directly proportional to the impurity areal density $n_{\text{imp}}$ and the Fermi velocity $v_F$. Thus, the elastic momentum relaxation rate is given by $1/\tau_{\text{tr}}(k) = n_{\text{imp}} v_F \sigma_{\text{tr}}(k)$, while the non-Hermitian local carrier loss or injection rate is $\gamma_{\text{nh}}(k) = n_{\text{imp}} v_F \sigma_{\text{abs}}(k)$.
   
	Before formulating the transport equation using these rates, it is imperative to establish the regime of validity for the semiclassical approximation. The introduction of the imaginary part $W$ in the scatterer's potential induces a quantum spectral broadening in the quasiparticle states, characterized by $\Gamma \approx W$~\cite{kozii_Arx2017,Shen_PRL2018,Zyuzin_PRB2018,Jiang_APL2023}. For the quasiparticle picture to remain robust and to justify the separation of the collision integral into independent processes, we restrict our analysis to the weak non-Hermiticity limit, $W \ll U \ll \mathcal{E}_F$. Furthermore, we assume that this quantum broadening is overcome by the thermal broadening dictated by the reservoir, i.e., $\Gamma \ll k_B T$ \cite{Rammer_RMP1986,Haug_B2008}. Under this hierarchy of scales, the semiclassical wave-packet dynamics remain valid, and the impact of the complex potential can be treated phenomenologically through the Boltzmann equation.
	
	For a homogeneous system in a steady state subjected to a uniform electric field, the distribution function is written as $f_{\mathbf{k}} = f_{\mathbf{k}}^0 + \delta f_{\mathbf{k}}$, where $f_{\mathbf{k}}^0$ is the equilibrium Fermi-Dirac distribution. In the presence of non-Hermitian impurities, the linearized collision integral $I_{\text{col}}[\delta f_{\mathbf{k}}]$ naturally decomposes into two kinetic contributions of fundamentally different nature. The first is the standard elastic redistribution kernel, which relaxes the carrier momentum due to scattering anisotropy and is characterized by the transport time $\tau_{\text{tr}}(\mathbf{k})$ \cite{ziman}. The second contribution arises from the lack of flux conservation imposed by the imaginary potential, introducing the aforementioned rate $\gamma_{\text{nh}}(\mathbf{k})$.
	
	If the non-Hermitian rate were introduced natively, the bare collision term $I_{\text{nh}}^{\text{bare}}[f_{\mathbf{k}}] = -\gamma_{\text{nh}}(\mathbf{k}) f_{\mathbf{k}}$ would violate local particle conservation, precluding the existence of a homogeneous steady state and inducing macroscopic spatial gradients in the chemical potential ($\nabla \mu$). For the reference distribution $f_{\mathbf{k}}^0$ to remain stationary ($I_{\text{col}}[f_{\mathbf{k}}^0] = 0$), the graphene monolayer must be coupled to an external bath (reservoir) that continuously injects or drains carriers to perfectly compensate for the local defect-induced gain or loss.

    Experimentally, the required thermal reservoir can be realized through out-of-plane coupling architectures, whose role must perfectly counteract the non-Hermitian nature of the impurities to maintain steady-state equilibrium. In the loss regime ($W < 0$), where the complex scatterers continuously drain carriers from the transport channel, the reservoir must act as a global source; this can be achieved via uniform optical pumping \cite{Li_PRL2012} or continuous carrier injection from vertical electrical gates \cite{Britnell_Sci2012}. Conversely, in the gain regime ($W > 0$), where impurities act as local particle sources, the reservoir must act as a macroscopic sink to prevent population divergence. This mechanism emulates transverse leakage currents where excess electrons continuously tunnel into deep trap states within the substrate (e.g., amorphous SiO$_2$ \cite{Lee_APL2011}) or an adjacent dielectric layer \cite{Wang_ACSN2010}. With the inclusion of this reservoir, the non-Hermitian term acts exclusively on the deviation from equilibrium $\delta f_{\mathbf{k}}$, resulting in the effective collision integral:
	\begin{equation}
		I_{\text{col}}[\delta f_{\mathbf{k}}] = - \left[ \frac{1}{\tau_{\text{tr}}(\mathbf{k})} + \gamma_{\text{nh}}(\mathbf{k}) \right] \delta f_{\mathbf{k}}.
	\end{equation}
	Therefore, the relaxation of the system is governed by a modified, effective relaxation time $\tau_{\text{eff}}(\mathbf{k})$ defined as:
	\begin{equation}
		\frac{1}{\tau_{\text{eff}}(\mathbf{k})} = n_{\text{imp}} v_F \left[ \sigma_{\text{tr}}(\mathbf{k}) + \sigma_{\text{abs}}(\mathbf{k}) \right].
		\label{eq:tau_eff}
	\end{equation}
	As an additional note, it is worth emphasizing that in the strict perturbative limit addressed in this work ($W \ll U$), the non-Hermitian rate $\gamma_{\text{nh}}$ acts as a small thermodynamic correction to the elastic transport time, yielding $\tau_{\text{eff}} \approx \tau_{\text{tr}} - \tau_{\text{tr}}^2 \gamma_{\text{nh}}$. From a kinetic standpoint, this reveals a profound symmetry between dissipation and amplification. Absorption introduces an additional relaxation channel that shortens the effective lifetime, in strict structural accordance with Matthiessen's rule for independent scattering rates \cite{ziman}. Conversely, weak gain acts as a 'negative' scattering mechanism that replenishes forward momentum, thereby increasing the effective lifetime. However, this enhancing is bounded by the macroscopic stability condition $1/\tau_{\text{eff}} > 0$. In such case, the restoring force of collisions is inverted: infinitesimal fluctuations are exponentially amplified rather than damped.

\subsection{Thermoelectric Coefficients}
\label{subsec:Thermoelectric}
	
	Within the relaxation-time approximation, the linear response of the electron gas to external electric fields and thermal gradients is elegantly captured by the Onsager matrix formalism. For a system driven by an electrochemical field $\mathbf{E} - (1/q)\bm{\nabla}\mu$ and a temperature gradient $\bm{\nabla} T$, the charge current $\mathbf{J}^1$ and the heat current $\mathbf{J}^2$ are defined through the transport coefficients $L^{(ij)}$ as
	\begin{align}
		\mathbf{J}^1 &= \frac{q^2}{T} L^{(11)} \left( \mathbf{E} - \frac{1}{q}\bm{\nabla}\mu \right) + \frac{q}{T^2} L^{(12)} (-\bm{\nabla} T), \\[6pt]
		\mathbf{J}^2 &= \frac{q}{T} L^{(21)} \left( \mathbf{E} - \frac{1}{q}\bm{\nabla}\mu \right) + \frac{1}{T^2} L^{(22)} (-\bm{\nabla} T),
	\end{align}
	where $q = -e$ is the electron charge. For monolayer graphene in the conduction band ($E = \hbar v_F k > 0$), accounting for the spin and valley degeneracies ($g_s g_v = 4$), the isotropic Onsager coefficients take the form:
	\begin{equation}
		L^{(ij)}(\mu, T) = 2 v_F^2 T \int_0^\infty dE \, D(E) \tau_{\text{eff}}(E) (E - \mu)^{i+j-2} \left( -\frac{\partial f^0}{\partial E} \right).
		\label{eq:onsager}
	\end{equation}
	Here, $D(E) = E / (2\pi \hbar^2 v_F^2)$ is the density of states per unit area, per spin, and per valley. 
	
	In standard Hermitian transport theory, the low-temperature limit ($k_B T \ll \mathcal{E}_F$) typically allows for the analytical evaluation of these integrals via the Sommerfeld expansion. However, the introduction of the complex potential $V_0 = U + iW$ induces a strong energy dependence in both the partial-wave phase shifts and the non-Hermitian cross-section $\sigma_{\text{abs}}(E)$. Furthermore, the finite quantum broadening $\Gamma$ invalidates the assumption that the thermal window is the sole source of smearing around the Fermi level. Consequently, the Sommerfeld expansion breaks down, and the integrals in Eq.~\eqref{eq:onsager} must be evaluated exactly over the entire thermal window using the full energy-dependent effective time $\tau_{\text{eff}}(E)$. 
	
	From the Onsager matrix elements, the macroscopic observable quantities can be directly extracted. The DC electrical conductivity under isothermal conditions ($\bm{\nabla} T = 0$) is given by:
	\begin{equation}
		\sigma^{\text{DC}} = \frac{e^2}{T} L^{(11)}.
        \label{eq:DC_cond}
	\end{equation}
	The electronic thermal conductivity $\kappa^{\text{el}}$, defined as the thermal response under open-circuit conditions ($\mathbf{J}^1 = 0$), is expressed as:
	\begin{equation}
		\kappa^{\text{el}} = \frac{1}{T^2} \left[ L^{(22)} - \frac{\left( L^{(12)} \right)^2}{L^{(11)}} \right].
	\end{equation}
	Similarly, the Seebeck coefficient (or thermopower) $S$, which measures the voltage generated in response to a temperature gradient at zero charge current, is computed as:
	\begin{equation}
		S = \frac{1}{qT} \frac{L^{(12)}}{L^{(11)}}.
	\end{equation}
	Finally, to evaluate the maximum efficiency of the electronic energy conversion, we define the purely electronic thermoelectric figure of merit:
	\begin{equation}
		ZT_{\text{el}} = \frac{S^2 \sigma^{\text{DC}} T}{\kappa^{\text{el}}}.
	\end{equation}
	
	By exploiting the weak non-Hermiticity limit ($W \ll U$), one can substitute the linear expansion of the effective relaxation time, $\tau_{\text{eff}} \approx \tau_{\text{tr}} - \tau_{\text{tr}}^2 \gamma_{\text{nh}}$, directly into Eq.~\eqref{eq:onsager}. This procedure decouples each transport coefficient into a purely Hermitian baseline contribution plus a first-order thermodynamic correction scaling linearly with the loss/gain parameter $W$, elegantly elucidating how non-unitary scattering mechanisms systematically degrade or enhance macroscopic thermoelectric responses.

\section{Results and Discussion} 
\label{sec:results_discussion}

In this section, we perform the numerical evaluation of the macroscopic electronic and thermoelectric quantities. Following previous works \cite{Dean2010, CBM_PB2026}, we consider non-Hermitian circular impurities with a radius of $a = 3$ nm, randomly distributed with density  $n_{\text{imp}} = 10^{12} \,\text{cm}^{-2}$. We restrict our analysis to the electron-doped regime ($s\chi = 1$), placing the Fermi level in the conduction band. Standard values for the Fermi energy in electrostatically gated graphene typically lie in the range of $\mathcal{E}_F \approx 200$ meV \cite{DasSarmaRMP2011}. However, under the continuous carrier injection or optical pumping mechanisms required to sustain the macroscopic steady-state in the gain regime ($W > 0$), the quasi-Fermi level can be driven up to $\sim 400$ meV \cite{Li_PRL2012}. Throughout these ranges, the continuum linear Dirac model remains a highly accurate effective description, as band structure corrections only become significant at energies approaching $1$ eV.

    The scattering effects of circular impurities, modeled by either constant or oscillator-like potentials, have been extensively discussed in previous works \cite{CBM_PB2026, CBM_R2025}. In particular, the isolated influence of strictly real potentials across a wide range of scatterer sizes was comprehensively analyzed in \cite{CBM_PB2026}. For this reason, in the present work we limit our study to isolating the novel effects introduced by the non-Hermitian gain/loss parameter $W$. To this end, we fix both the impurity radius ($a=3$ nm) and the Hermitian potential depth ($U=20$ meV) at representative values. We emphasize that, to ensure the validity of the semiclassical steady-state framework, our analysis is strictly confined to the weak non-Hermiticity limit, $\vert{}W\vert{} \ll U$.

\subsection{Scattering cross-sections}

\begin{figure}
    \centering
    \begin{subfigure}{0.48\textwidth}
        \includegraphics[width=\linewidth]{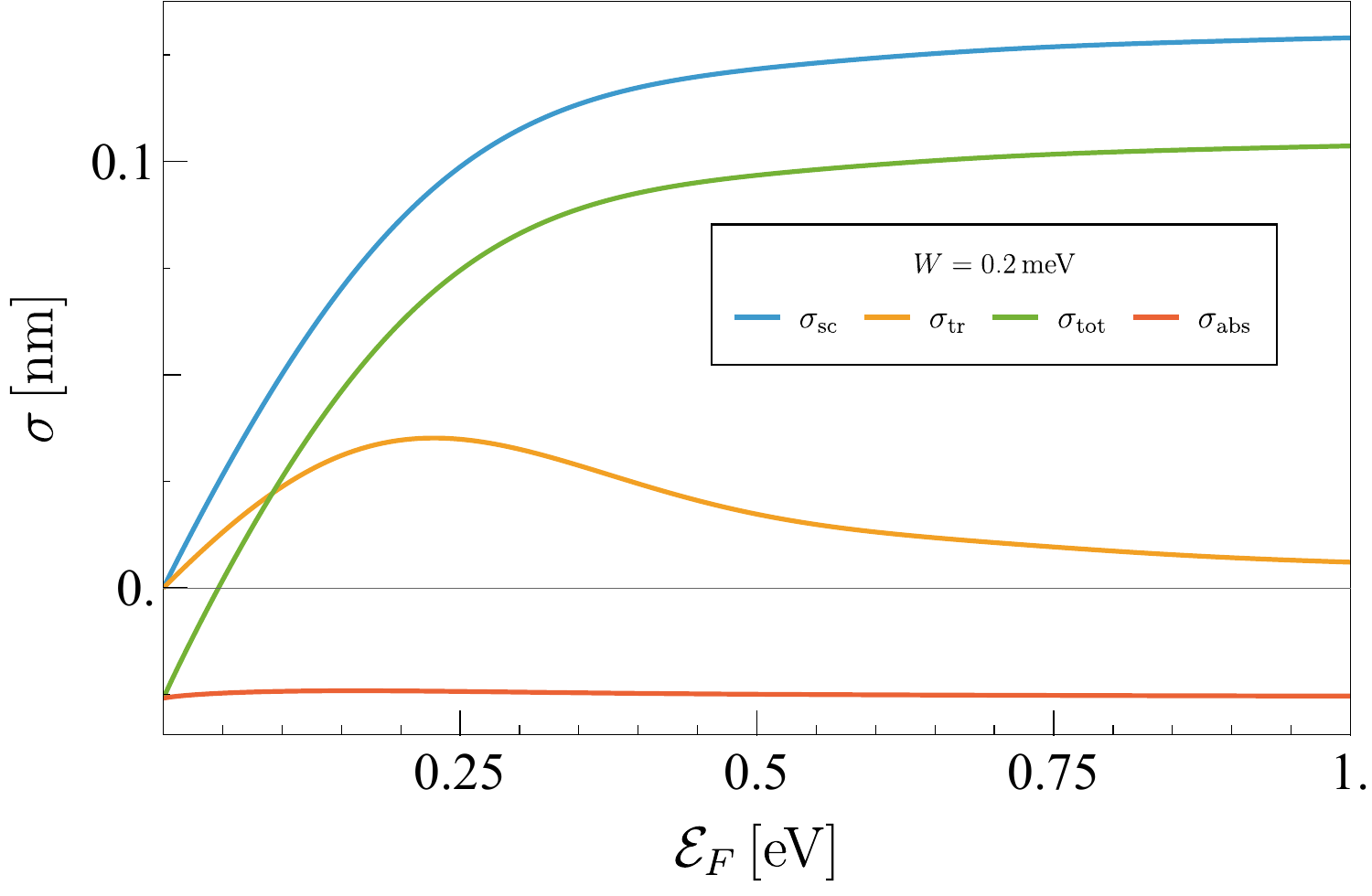}
        \caption{}
        \label{Fig:CrossS_p}
    \end{subfigure}
    \begin{subfigure}{0.48\textwidth}
        \includegraphics[width=\linewidth]{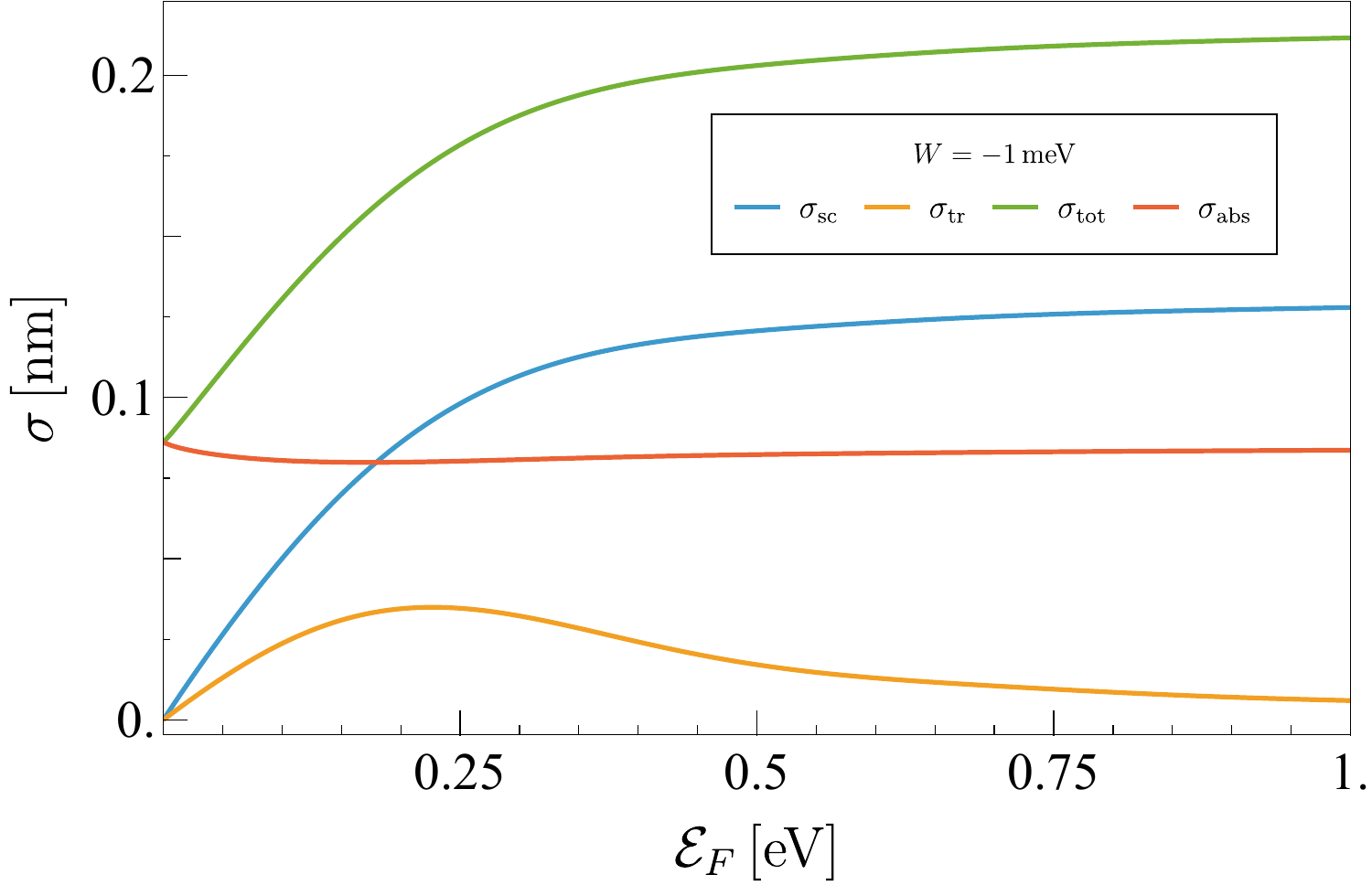}
        \caption{}
        \label{Fig:CrossS_n}
    \end{subfigure}
    \caption{Energy dependence of the macroscopic scattering cross-sections for non-Hermitian impurities with radius $a = 3$ nm and Hermitian depth $U = 20$ meV. Top panel (a): Active source regime ($W = 0.2$ meV). The non-Hermitian cross-section ($\sigma_{\text{abs}}$) is strictly negative. At low energies, the total cross-section ($\sigma_{\text{tot}}$) drops below zero, indicating the threshold where active gain overcomes elastic scattering and destabilizes the semiclassical Boltzmann steady-state. Bottom panel (b): Absorptive sink regime ($W = -1$ meV). The strictly positive $\sigma_{\text{abs}}$ heavily dominates the scattering landscape at low energies, drastically increasing the overall cross-section and suppressing transport. Across both panels, the pronounced geometric reduction of the transport cross-section ($\sigma_{\text{tr}}$) relative to the elastic one ($\sigma_{\text{sc}}$) at higher energies reflects the chiral nature of Dirac fermions.}
    \label{Fig:Sc_Cross}
\end{figure}

    To gain deeper microscopic insight into the collision rates governing the macroscopic steady-state, it is instructive to first analyze the energy dependence of the individual scattering cross-sections. Figure \ref{Fig:Sc_Cross} presents a comparative evaluation of the elastic ($\sigma_{\text{sc}}$), transport ($\sigma_{\text{tr}}$), absorption ($\sigma_{\text{abs}}$), and total ($\sigma_{\text{tot}}$) cross-sections for both active source ($W = 0.2$ meV, Fig \ref{Fig:CrossS_p}) and absorptive sink ($W = -1$ meV, Fig \ref{Fig:CrossS_n}) impurities across the conduction band.

    An standard feature across both regimes is the stark geometric reduction of the transport cross-section $\sigma_{\text{tr}}$ relative to the purely elastic cross-section $\sigma_{\text{sc}}$. While $\sigma_{\text{sc}}$ grows and largely saturates at higher incident energies, $\sigma_{\text{tr}}$ peaks at low energies and subsequently undergoes a pronounced suppression. This discrepancy fundamentally arises from the chiral nature of massless Dirac fermions \cite{CastroNetoRMP2009, NovikovPRB2007}. At higher incident energies ($ka \gtrsim 1$), the elastic scattering amplitude becomes highly anisotropic and strongly forward-peaked, a known manifestation of Klein tunneling dynamics in finite-range potentials \cite{WuPRB2014}. Because the transport cross-section weighs scattering events by the geometric factor $(1 - \cos\theta)$, this dominant forward scattering is heavily penalized, reflecting the physical reality that small-angle deflections do not efficiently degrade longitudinal momentum along the transport channel \cite{DasSarmaRMP2011}.  

    Beyond the purely elastic dynamics, the complex potential introduces profound macroscopic consequences at low energies. For an active source (Fig. \ref{Fig:CrossS_p}), the absorption cross-section $\sigma_{\text{abs}}$ remains strictly negative, driving the continuous local injection of carrier flux. Strikingly, as the incident energy approaches the Dirac point, the elastic cross-sections vanish, whereas the non-Hermitian cross-section converges to a finite, non-zero value \cite{NovikovPRB2007, CBM_PB2026}. Consequently, the total cross-section $\sigma_{\text{tot}}$ (and critically, the effective transport kinetic sum $\sigma_{\text{tr}} + \sigma_{\text{abs}}$) becomes negative. Physically, this marks the stability threshold where active carrier gain completely overwhelms the elastic angular relaxation. In this regime, the restoring force of collisions is inverted ($1/\tau_{\text{eff}} < 0$), causing the linearized Boltzmann steady-state solution to become dynamically unstable and strictly bounding the validity of the semiclassical response theory

    Conversely, for an absorptive sink (Fig. \ref{Fig:CrossS_n}), the non-Hermitian cross-section is positive and dominates the entire scattering landscape at low energies. By adding a massive, momentum-independent inelastic sink to the scattering events, the localized loss drastically increases the effective cross-section. This mechanism thoroughly suppresses the macroscopic transport time, heavily degrading carrier propagation long before standard elastic momentum redistribution can occur.

\subsection{Transport relaxation time}

\begin{figure}
        \centering
        \begin{subfigure}
            {0.48\textwidth}
            \includegraphics[width=\linewidth]{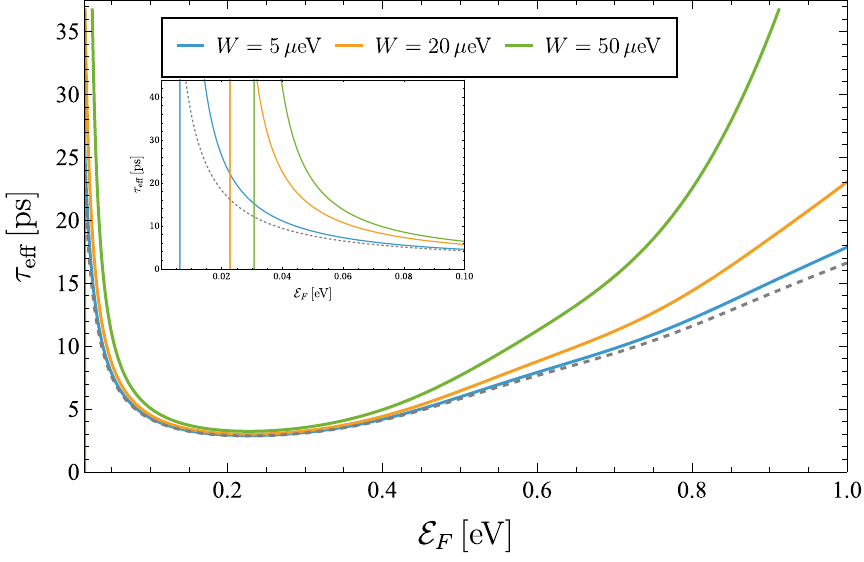}
            \subcaption{}
            \label{Fig:TauvsEf_p}
        \end{subfigure}
        \begin{subfigure}
            {0.48\textwidth}
            \includegraphics[width=\linewidth]{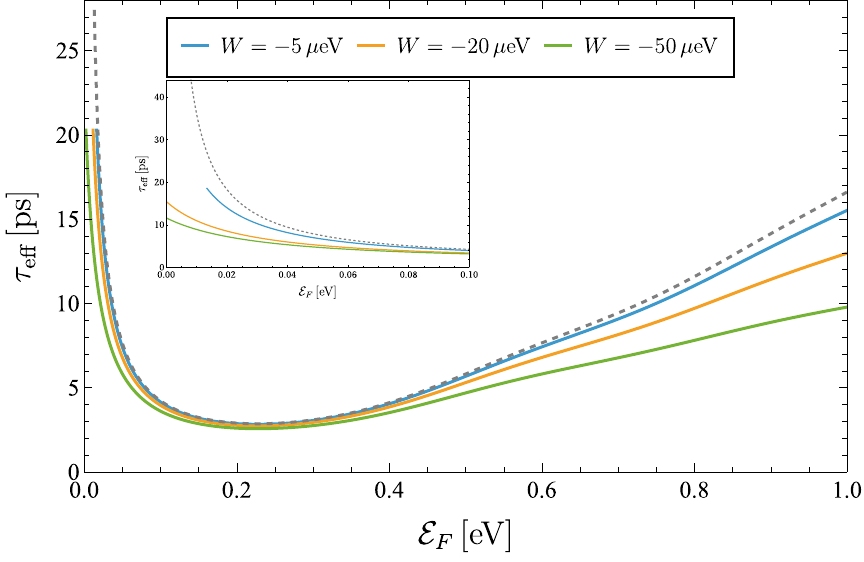}
            \subcaption{}
            \label{fig:TauvsEf_n}
        \end{subfigure}
        \caption{Plots of the effective transport relaxation time, calculated using \eqref{eq:tau_eff}, for several positive (top) and negative (bottom) values of the non-Hermiticity parameter $W$. The dotted gray line indicates the Hermitian baseline ($W=0$).}
        \label{fig:TauvsEf}
    \end{figure}

    Figure \ref{fig:TauvsEf} illustrates the energy dependence of the effective relaxation time for various positive (top) and negative (bottom) values of the non-Hermiticity parameter $W$. These results graphically reinforce the kinetic framework established at the end of Sec. \ref{subsec:EffectiveTime}. As seen in the bottom panel (Fig. \ref{fig:TauvsEf}b), absorptive impurities ($W < 0$) act as particle sinks, opening an inelastic-like relaxation channel that increases the total scattering rate and systematically reduces $\tau_{\text{eff}}$ across the entire spectrum. Conversely, the top panel demonstrates how weak gain ($W > 0$) acts as a local source that replenishes forward momentum, reducing the net scattering rate and thereby enhancing the effective carrier lifetime.
    
    Furthermore, a striking and unanticipated feature emerges in the low-energy limit, strictly for the loss regime. Although it is well established that the presence of impurities in graphene enhances the density of states at the neutrality point \cite{Peres_PRB2006}, standard semiclassical theory fails to properly describe this behavior, yielding an unphysical divergence of the elastic transport time as $\mathcal{E}_F \to 0$ (as seen in Fig. \ref{Fig:TauvsEf_p} and extensively discussed in previous works \cite{CBM_PB2026,CBM_R2025,Auslender_PRB2007}). However, as observed in the leftmost region of Fig. \ref{fig:TauvsEf_n}, the inclusion of absorptive impurities ($W < 0$) effectively suppresses this divergence, suggesting that the non-Hermitian term naturally captures this impurity-induced renormalization of the density of states. From a physical standpoint, slow-moving carriers near the neutrality point are highly susceptible to being captured by the localized sinks. As the energy approaches zero, this probability of absorption drastically increases, overwhelmingly dominating the transport dynamics. Consequently, carriers are depleted from the channel long before they can undergo standard elastic collisions, forcing the effective relaxation time to stabilize. This implies that in the immediate vicinity of the Dirac point, macroscopic transport is governed entirely by non-Hermitian flux depletion rather than by elastic momentum redistribution, marking a regime where dissipation fundamentally overrides standard quasiparticle dynamics. 

\subsection{Electrical and thermal conductivities}
    In this subsection, we evaluate the DC electrical ($\sigma$) and electronic thermal ($\kappa$) conductivities. Because the scattering potential is isotropic and the system is assumed to be homogeneous, the transport tensors are isotropic and diagonal, allowing us to restrict the analysis to the longitudinal components of the tensors. We note that within the relaxation-time approximation, these conductivities scale inversely with the impurity concentration $n_{\text{imp}}$. Consequently, modulating the defect density within the bounds of the dilute limit provides a straightforward experimental knob for fine-tuning the magnitude of the macroscopic transport coefficients.

    \begin{figure}
        \centering
        \begin{subfigure}
            {0.48\textwidth}
            \includegraphics[width=\linewidth]{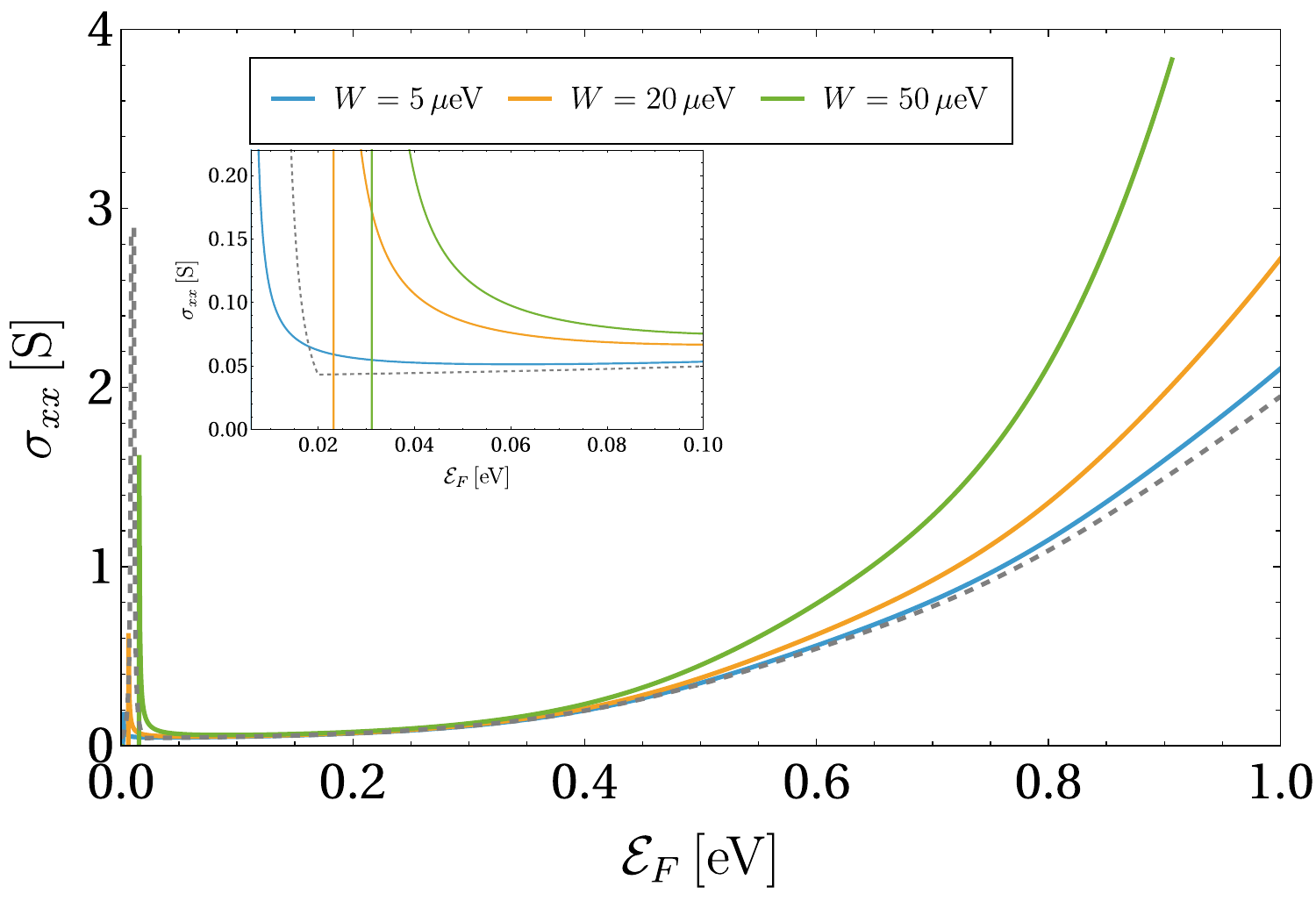}
            \subcaption{}
            \label{Fig:SigmavsEf_p}
        \end{subfigure}
        \begin{subfigure}
            {0.48\textwidth}
            \includegraphics[width=\linewidth]{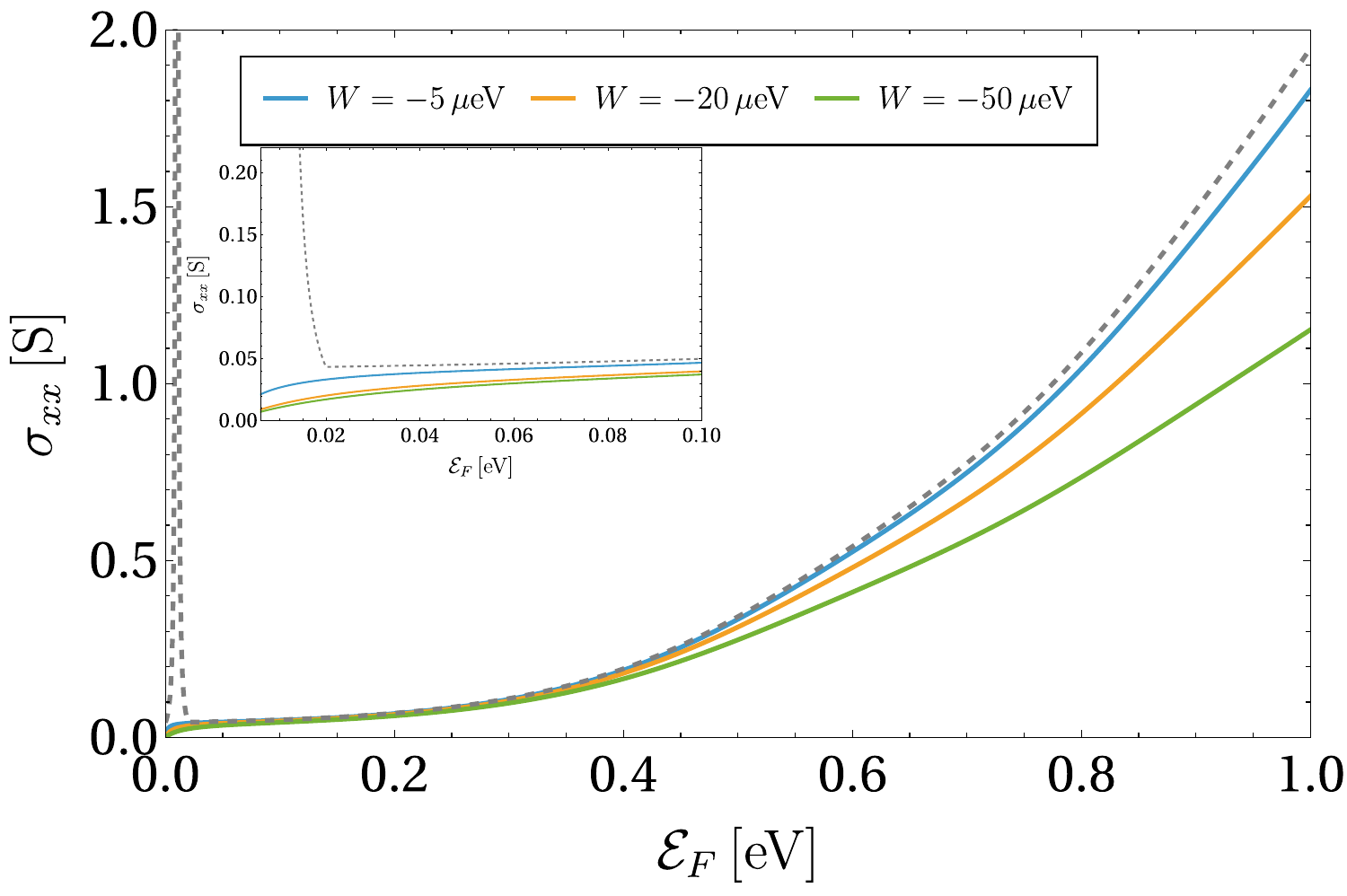}
            \subcaption{}
            \label{fig:SigmavsEf_n}
        \end{subfigure}
        \caption{Plots of the longitudinal component of the DC electrical conductivity, calculated using \eqref{eq:DC_cond}, for several positive (top) and negative (bottom) values of the non-Hermiticity parameter $W$. The dotted gray line indicates the Hermitian baseline ($W=0$).}
        \label{fig:SigmavsEf}
    \end{figure}
\subsubsection{DC electrical conductivity}
    Figure \ref{fig:SigmavsEf} displays the residual DC electrical conductivity, computed from Eq. \eqref{eq:DC_cond} in the zero-temperature limit ($T \to 0$). As dictated by the Onsager formalism, the energy dependence of the zero-temperature conductivity is essentially governed by the effective relaxation time weighted by the density of states, scaling as $\sigma \propto \mathcal{E}_F \tau_{\text{eff}}(\mathcal{E}_F)$. Previous works have shown that, for Hermitian impurities, this linear energy weighting is insufficient to overcome the strong low-energy divergence of the transport time, ultimately yielding an unphysical divergence in the residual conductivity as $\mathcal{E}_F \to 0$ \cite{CBM_PB2026,CBM_R2025}. Remarkably, as revealed by the plots, the introduction of the complex potential cures this anomaly across both the loss and gain regimes, albeit exhibiting distinct low-energy profiles. In the absorptive case ($W < 0$), the loss rate forces the effective transport time to a finite value at low energies, leading to the suppression of the residual conductivity near the Dirac point. Conversely, in the active source regime ($W > 0$), the zero-energy divergence is similarly circumvented; however, rather than a flat suppression, the interplay between the vanishing density of states and the gain-induced scattering modifications manifests as small, finite peaks in the immediate vicinity of $\mathcal{E}_F = 0$. The amplitude of these non-divergent peaks is directly correlated with the magnitude of $W$. Although this mathematically induced regularization does not fully capture the universal minimum conductivity plateau observed experimentally \cite{NovoselovNature2005,DasSarmaRMP2011}, it is nonetheless sufficient to regularize the kinetic kernel $\sim k \tau_{\text{eff}}(k)$ in the integrand of Eq. \eqref{eq:onsager}. This stabilization maps the macroscopic conductivity to well-behaved, finite values, successfully overcoming the inherent breakdown of the standard semiclassical Boltzmann theory at the neutrality point.

\begin{figure}
        \centering
        \begin{subfigure}{0.48\textwidth}
            \includegraphics[width=\linewidth]{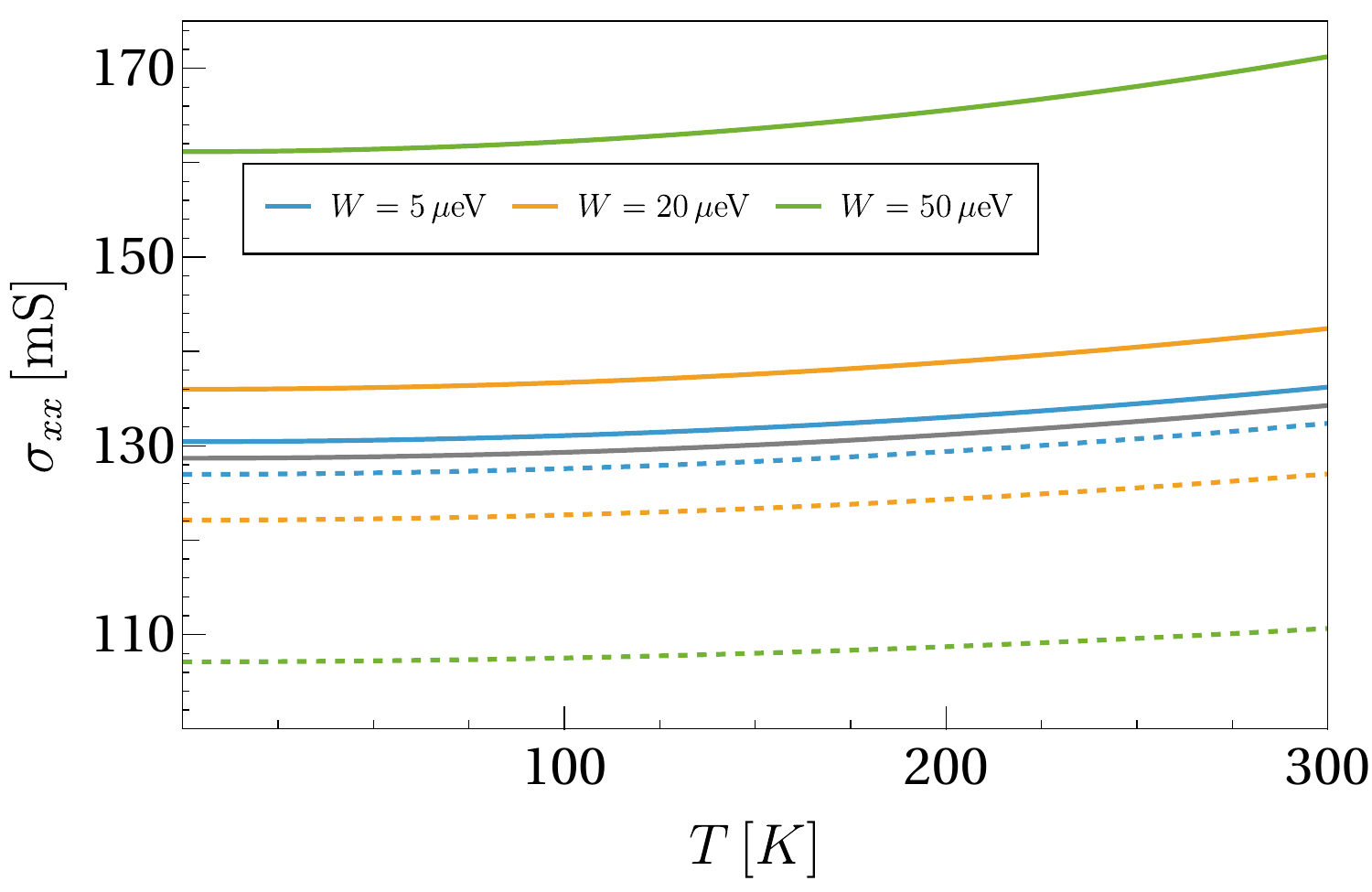}
            \caption{}
            \label{fig:SigmavsT}
        \end{subfigure}
        \begin{subfigure}{0.48\textwidth}
            \includegraphics[width=\linewidth]{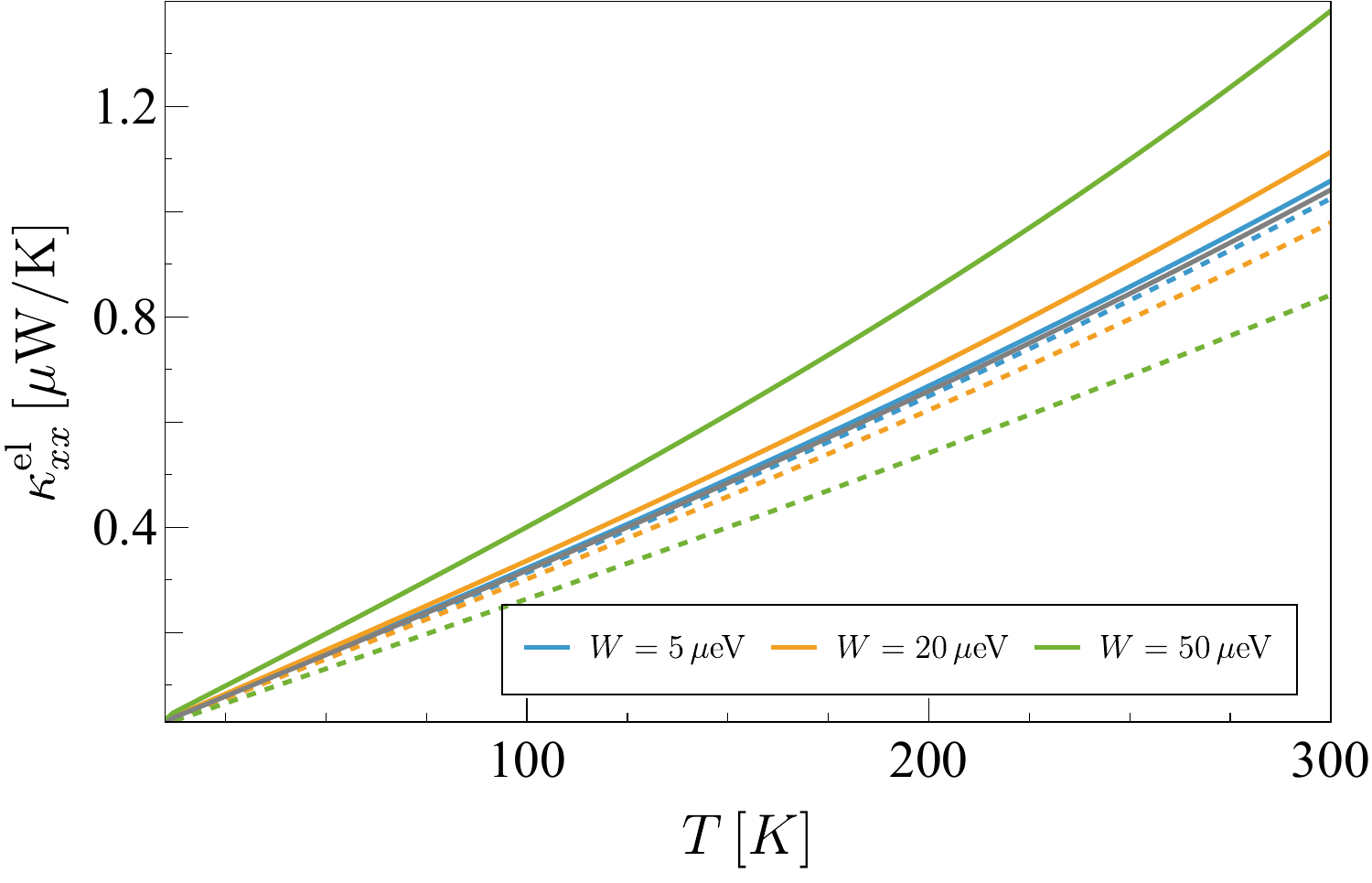}
            \caption{}
            \label{fig:KapvsT}
        \end{subfigure}
        \begin{subfigure}{0.48\textwidth}
            \includegraphics[width=\linewidth]{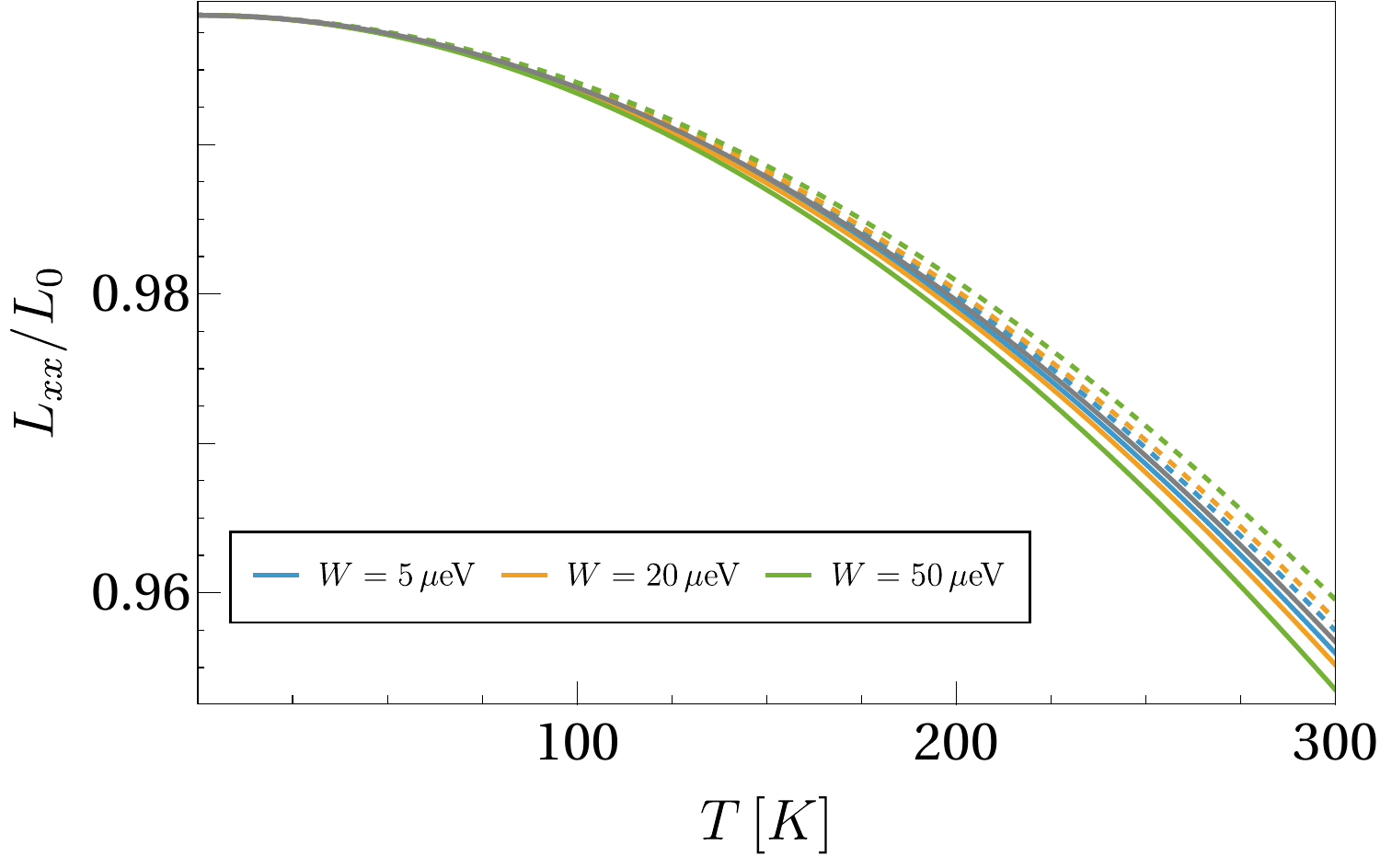}
            \caption{}
            \label{fig:LorenzvsT}
        \end{subfigure}
        \caption{Temperature dependence of the transport coefficients at $\mathcal{E_F}=330$ meV for several non-Hermitian potential strengths $W$. Top panel (a): longitudinal DC electrical conductivity $\sigma(T)$. Central panel (b): electronic thermal conductivity $\kappa^{\text{el}}(T)$. Bottom panel (c): Normalized Lorenz number $L/L_0$. Across all panels, Solid and dotted lines correspond to gain ($W>0$) and loss ($W<0$) regimes, respectively, while the solid gray line is the strictly Hermitian baseline $W=0$. All curves are computed using the exact integral representation of the Onsager coefficients given in Eq. \eqref{eq:onsager}.}
        \label{fig:CondvsT}
\end{figure}

    To extend our analysis beyond the zero-temperature limit, we investigate the temperature dependence of the DC electrical conductivity. As a representative case for our thermal analysis, we fix the Fermi energy at $\mathcal{E}_F = 330$ meV. This highly doped regime is consistent with the continuous optical pumping or gating architectures (discussed in Sec. \ref{subsec:EffectiveTime}) required to sustain the macroscopic non-Hermitian steady state. Figure \ref{fig:SigmavsT} displays $\sigma$ as a function of temperature $T$ for various positive (solid lines) and negative (dotted lines) values of the non-Hermiticity parameter $W$. The conductivities were evaluated by numerically integrating the exact Onsager expression over the full thermal window, thus avoiding the conventional low-temperature approximations.
    
    A systematic comparison between our exact numerical integration and the results yielded by the standard Sommerfeld expansion reveals that, although the deviation between the approximate and exact results becomes more pronounced as the magnitude of $W$ increases, the discrepancy remains strictly of the same order of magnitude as the baseline error for the Hermitian limit (around $\sim 10^{-2}\,\mu$S). This implies that for the weak non-Hermiticity regime ($\vert{}W\vert{} \ll U$), the thermal smearing is not pathologically distorted by the complex potential. Consequently, the temperature profile of the conductivity is largely governed by the local derivative and curvature of the transport time at the Fermi level. At $\mathcal{E}_F = 330$ meV, the effective relaxation time varies smoothly, resulting in the relatively mild temperature dependence observed across all curves in Fig. \ref{fig:SigmavsT}. However, as can be inferred from the zero-temperature energy profiles in Fig. \ref{fig:SigmavsEf}, both the magnitude and the slope of $\tau_{\text{eff}}(\mathcal{E}_F)$ become highly sensitive to the non-Hermitian parameter at higher doping levels ($\mathcal{E}_F > 500$ meV). Therefore, while the thermal macroscopic deviations remain moderate at intermediate energies, pushing the system deeper into the high-energy regime would drastically amplify the differences between the gain, loss, and Hermitian thermal responses, driven by the steepening of the kinetic derivatives.
    
\subsubsection{Electronic thermal conductivity}

We now turn our attention to the evaluation of the electronic thermal conductivity $\kappa$. Figure \ref{fig:KapvsT} shows $\kappa$ as a function of temperature for various non-Hermitian parameters $W$, computed via the exact numerical integration of the Onsager coefficients defined in Eq. \eqref{eq:onsager}. At a macroscopic scale, the thermal conductivity exhibits the characteristic linear growth with temperature predicted by the standard Sommerfeld expansion. Consistent with our previous comparative analysis for the electrical response, the exact integral evaluation confirms that the Sommerfeld approximation remains highly accurate in this regime. While the magnitude of the thermal current is remarkably sensitive to the complex potential, increasing for the active source regime ($W > 0$) and decreasing for the absorptive sink regime ($W < 0$), the overall linear scaling remains structurally robust across the intermediate temperature range. Furthermore, it is worth noting that while recent studies on non-Hermitian potential barriers have demonstrated that strong active sources can drive a residual non-zero thermal conductivity even at absolute zero \cite{Bonilla_PRB2026}, our exact low-temperature evaluation confirms that such anomalous phenomena do not emerge in the perturbative quasi-Hermitian limit ($\vert{}W\vert{} \ll U$) considered here. Instead, the standard thermodynamic constraints of the Fermi gas are strictly preserved, ensuring that the macroscopic thermal conductivity inherently vanishes as $T \to 0$.

\subsubsection{Lorenz coefficient}

To complete our macroscopic analysis of the coupled charge and heat currents, we examine the temperature dependence of the normalized Lorenz number, $L/L_0$, where $L = \kappa^{\text{el}} / (T \sigma)$. The results are presented in the bottom panel of Fig. \ref{fig:CondvsT}. As expected from standard Fermi liquid theory, the results confirm the exact validity of the Wiedemann-Franz law in the extreme low-temperature limit ($T \to 0$), where $L/L_0 \to 1$. However, a clear deviation from ideal metallic behavior emerges as temperature increases. While the overall magnitude of this deviation is small, the complex potential subtly modulates the thermal response: weak gain ($W > 0$) systematically increases the deviation from the Wiedemann-Franz law, whereas absorptive loss ($W < 0$) suppresses it, keeping the system closer to the ideal Sommerfeld ratio. The relatively small magnitude of this overall deviation is consistent with our previous findings for purely Hermitian scatterers \cite{CBM_PB2026}, which demonstrated that the primary parameter governing the violation of the Wiedemann-Franz law is the spatial extent of the impurities. Since we have fixed the scatterer radius to $a=3$ nm, the system largely preserves the ideal metallic ratio, with the non-Hermitian parameter $W$ introducing only a fine-tuning effect over the baseline deviation.

Finally, it is instructive to place these results in the broader context of graphene transport literature. Experimental measurements close to the charge neutrality point have reported massive violations of the Wiedemann-Franz law, characterized by a pronounced enhancement of the Lorenz number \cite{Crossno2016}. Such observations, supported by theoretical frameworks incorporating bipolar diffusion \cite{Tu2023, PhysRevB.108.245415, ma14112704}, are widely interpreted as a signature of transport in a regime dominated by electron-hole coexistence and collective hydrodynamic effects. In contrast, the present work focuses on transport at a finite quasi-Fermi energy ($\mathcal{E}_F = 330$ meV), far from the neutrality point, where a single carrier type strictly dominates and hydrodynamic effects are strongly suppressed. Consequently, the deviations from the Sommerfeld value observed in Fig. \ref{fig:LorenzvsT} do not reflect collective bipolar phenomena, but rather emerge solely from the energy dependence of the non-Hermitian scattering rates at finite doping, consistent with the single-particle kinetic regime under consideration.

\subsection{Thermoelectric coefficients}

\begin{figure}
    \centering
    \begin{subfigure}{0.48\textwidth}
        \includegraphics[width=\linewidth]{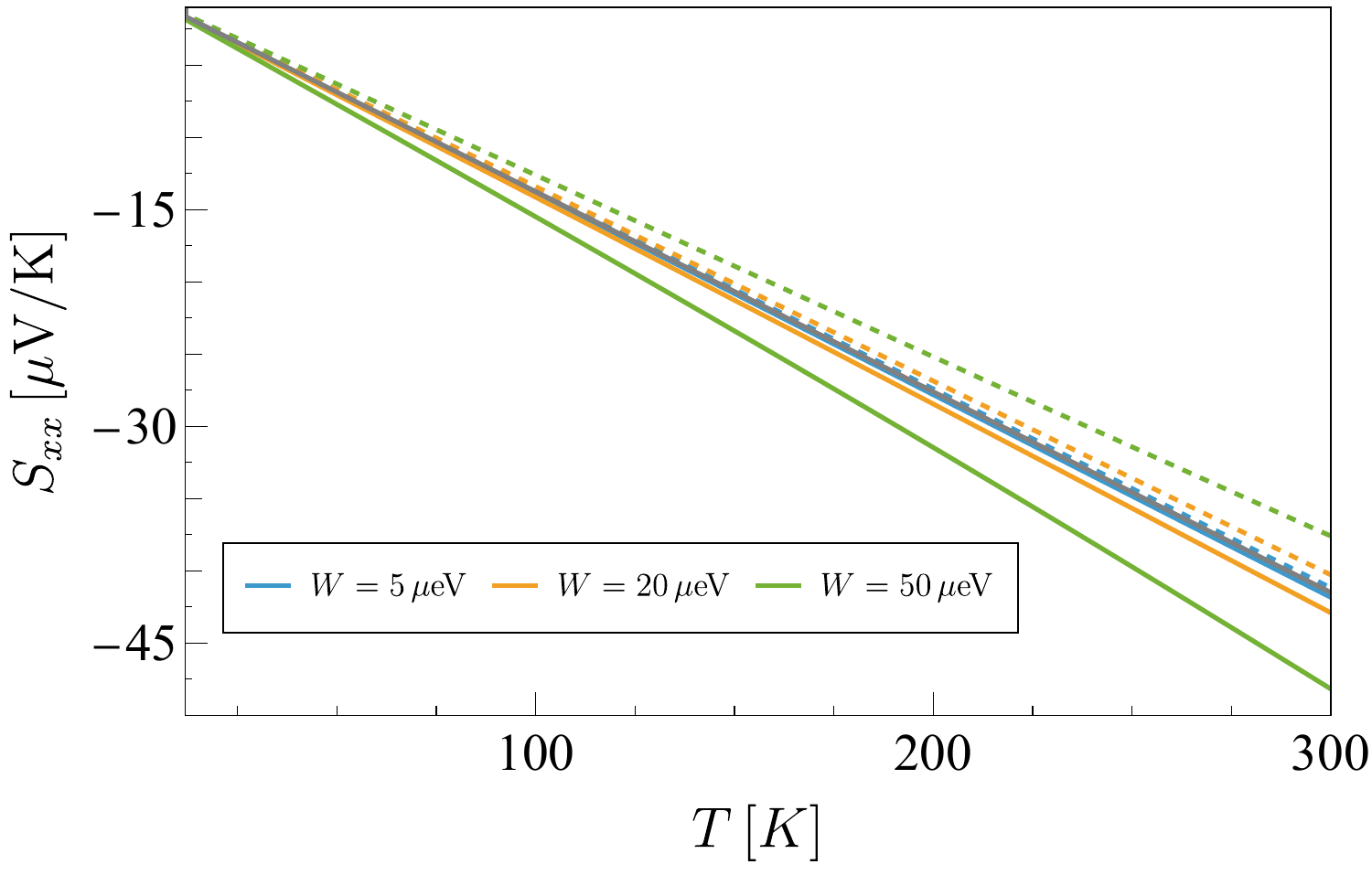}
        \caption{}
        \label{Fig:SeebeckvsT}
    \end{subfigure}
    \begin{subfigure}{0.48\textwidth}
        \includegraphics[width=\linewidth]{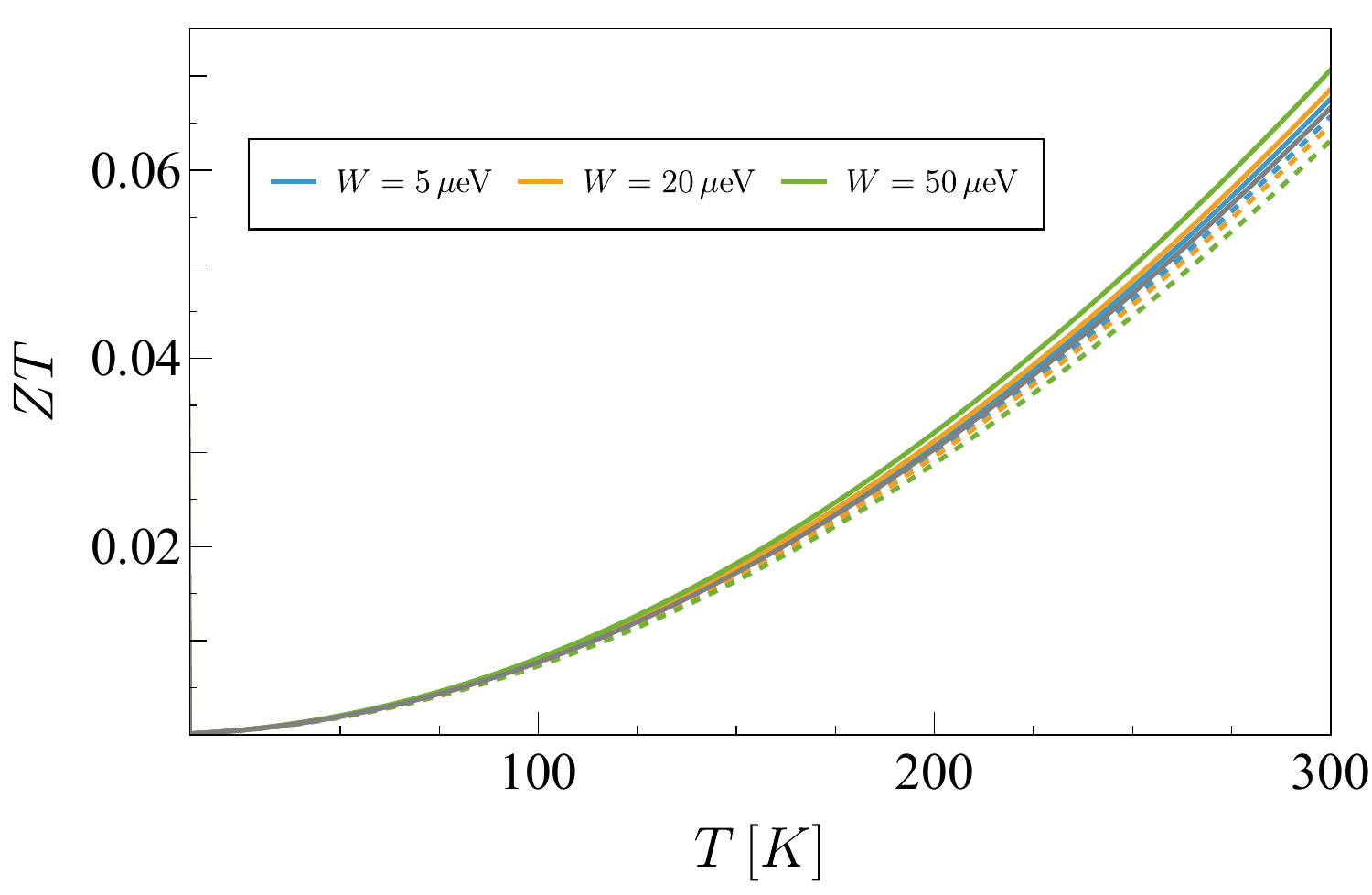}
        \caption{}
        \label{Fig:ZTvsT}
    \end{subfigure}
    \caption{Temperature dependence of the thermoelectric generation and efficiency at a fixed quasi-Fermi energy of $\mathcal{E}_F = 330$ meV. Top panel (a): Seebeck coefficient $S(T)$. Bottom panel (b): electronic thermoelectric figure of merit $ZT_{\text{el}}(T)$. Across both panels, solid and dotted lines correspond to the active source ($W > 0$) and absorptive sink ($W < 0$) regimes for various non-Hermitian potential strengths $W$, while the solid gray line indicates the strictly Hermitian baseline ($W = 0$). All curves are computed via the exact integral representation of the Onsager coefficients in Eq. \eqref{eq:onsager}.}
    \label{fig:ThermElec}
\end{figure}

We finally direct our attention to the thermoelectric generation capabilities of the non-Hermitian system. The Fig. \ref{Fig:SeebeckvsT} displays the temperature dependence of the Seebeck coefficient (or thermopower), $S(T)$. The strictly negative sign of $S$ across the entire temperature range confirms that transport is heavily dominated by electron-like quasiparticles in the conduction band. Consistent with standard diffusive transport, the magnitude of the thermopower grows approximately linearly with temperature \cite{CBM_R2025,CBM_PB2026,Hwang_PRB2009}. Remarkably, the inclusion of the complex potential introduces a systematic modulation of the Seebeck response: active sources ($W > 0$) significantly enhance the magnitude of the thermopower, whereas localized absorptive sinks ($W < 0$) degrade it. As dictated by standard semiclassical transport theory \cite{ziman}, the Seebeck coefficient is intimately linked to the energy asymmetry of the effective scattering time near the Fermi level. The gain-induced enhancement of the carrier lifetime effectively steepens the kinetic derivatives at higher energies, thereby boosting the thermodynamic driving force for the charge carriers.

It is instructive to contextualize this non-Hermitian modulation against the strictly Hermitian baseline ($W=0$). As discussed in previous works on finite-range disorder \cite{CBM_PB2026}, and as in the case of the Lorenz number, the purely Hermitian thermopower is highly sensitive to the spatial extent of the scatterers, while the potential depth $U$ has a nearly negligible impact. Our baseline generates Seebeck values whose magnitude is consistent with the $40\text{--}90\ \mu\text{V/K}$ range reported experimentally for doped graphene systems \cite{HWANG2023467}. The non-Hermitian parameter $W$ thus emerges as a novel and independent tuning knob, capable of significantly amplifying or suppressing the thermopower without altering the physical footprint of the impurities.

To contextualize the practical implications of these results, it is essential to evaluate the magnitude of the figure of merit. For practical applications, a total efficiency of $ZT \sim 1$ is typically required, while values exceeding $2$ are considered excellent. In our current model, the maximum values obtained for the purely electronic figure of merit reach $ZT_{\text{el}} \sim 0.1$. This result is consistent with the lower range of values reported for graphene nanoribbons \cite{Bonilla_PRB2026,ApplPhysA2024a,ZT_1}. In strictly Hermitian systems, the strength of the scattering potential has a relatively weak influence on $ZT$, leaving the radius of the scatterers as the dominant tuning parameter. However, the introduction of the non-Hermitian potential overcomes this geometric limitation. As demonstrated in Fig. \ref{fig:ThermElec}b, active gain ($W > 0$) provides a robust, independent mechanism to enhance the electronic thermoelectric efficiency without requiring larger structural defects.  

Interestingly, this macroscopic enhancement contrasts sharply with recent findings in the coherent quantum transport regime. In a recent study evaluating transport through a finite-size non-Hermitian potential barrier in a graphene nanoribbon using the Landauer formalism \cite{Bonilla_PRB2026}, the opposite trend was observed: the thermoelectric figure of merit was enhanced by absorptive loss ($W < 0$) and suppressed by active gain ($W > 0$). This dichotomy highlights the fundamental physical difference between coherent transmission across a localized non-Hermitian barrier and macroscopic diffusive transport through a dilute ensemble of non-Hermitian scatterers. In the semiclassical Boltzmann framework considered here, the thermoelectric response is fundamentally governed by the energy asymmetry of the macroscopic relaxation time, where the gain-induced increase in the carrier lifetime actively boosts the thermopower.

Although the isolated electronic values presented here do not represent a definitive solution to the low thermoelectric efficiency of macroscopic graphene, they present a clear pathway for improvement. In real devices, the total figure of merit is heavily suppressed by the massive phononic thermal conductivity of the lattice, $\kappa_{\text{ph}}$. Because the non-Hermitian impurities studied here primarily manipulate electronic transport, strategies that target the reduction of $\kappa_{\text{ph}}$, such as the chevron-type geometry and isotope engineering shown to achieve $ZT \sim 2.45$ in graphene \cite{ZT_2}, could be synergistically combined with impurity engineering to achieve high efficiency. Ultimately, this demonstrates that complex, non-unitary scattering potentials offer a powerful new degree of freedom to maximize the electronic contribution to graphene's thermoelectric performance.

\section{Conclusions} 
\label{sec:conclusions}

We have developed a semiclassical framework to describe charge and thermoelectric transport in electron-doped graphene containing a dilute distribution of finite-range non-Hermitian scattering centers. Starting from the exact partial-wave solution of the Dirac equation for a complex circular potential, we obtained the nonunitary scattering matrix and identified two distinct microscopic contributions to transport: elastic momentum relaxation, governed by the transport cross section, and carrier loss or gain, encoded in the absorption cross section. Within a stationary Boltzmann description sustained by an external reservoir, these two mechanisms combine into an effective relaxation rate that provides the link between local non-Hermitian scattering and macroscopic transport.

The non-Hermitian contribution produces a clear asymmetry between absorptive and amplifying impurities. Localized loss introduces an additional relaxation channel that increases the total scattering rate and shortens the effective carrier lifetime, whereas weak gain reduces the net relaxation rate and enhances the lifetime, subject to the stability condition $1/\tau_{\rm eff}>0$. At low energies, the absorption cross section becomes particularly important and can dominate over elastic momentum relaxation. In the loss regime this suppresses the divergence of the semiclassical relaxation time near the Dirac point, while in the gain regime the approach to the instability threshold places a natural bound on the validity of the steady-state Boltzmann description.

These microscopic changes are directly reflected in the macroscopic transport coefficients. Weak gain enhances both the electrical and electronic thermal conductivities, while absorptive loss suppresses them. Nevertheless, the Wiedemann--Franz relation remains robust in the low-temperature limit, with the Lorenz ratio approaching the Sommerfeld value as $T\rightarrow 0$. At finite temperature, non-Hermiticity mainly modifies the small deviations from this limit through the energy dependence of the effective relaxation time, rather than generating a qualitatively different thermal transport regime in the weakly non-Hermitian limit considered here.

The most significant effect appears in the thermoelectric response. Active gain increases the magnitude of the Seebeck coefficient by enhancing the energy asymmetry of the transport kernel around the Fermi level and, consequently, increases the electronic figure of merit $ZT_{\rm el}$. Absorptive loss produces the opposite trend. This result is particularly noteworthy because it contrasts with the behavior previously reported for coherent transport through a finite non-Hermitian barrier in graphene, where loss enhances the thermoelectric figure of merit and gain suppresses it. The comparison shows that the thermoelectric consequences of non-Hermiticity cannot be characterized by gain or loss alone, but depend crucially on the transport regime and on whether the relevant physics is governed by coherent transmission or by the energy dependence of a diffusive relaxation time.

Our results therefore identify non-Hermitian scattering as an additional degree of freedom for tailoring thermoelectric transport in graphene, complementary to conventional control through impurity density, size, and electrostatic strength. The present analysis is restricted to dilute disorder, weak non-Hermiticity, finite electron doping, and the electronic contribution to thermal transport. A complete assessment of thermoelectric performance should also incorporate the lattice thermal conductivity and the energetic cost of sustaining the external reservoir required to maintain the stationary state.  Extending the present framework toward stronger non-Hermiticity, multiple-scattering effects, and fully open-system transport descriptions may clarify how far the gain-induced enhancement found here can be pushed and whether analogous mechanisms can be realized in other Dirac and topological materials.

\section*{Acknowledgments}
	J.A.C. gratefully acknowledges the support of SECIHTI through the program \textit{Becas Nacionales para estudios de Posgrado}, under grant number 4018746. D.A.B. was supported by the DGAPA-UNAM Posdoctoral Program. A.M.-R. acknowledges financial support by UNAM-PAPIIT project No. IG100224, UNAM-PAPIME project No. PE109226, by SECIHTI project No. CBF-2025-I-1862 and by the Marcos Moshinsky Foundation.
	
	
	\bibliography{cas-refs}

@article{CC_Kim_2025,
    author = {Kim, Jewook and Chang, Hwanseok and Bae, Gwangmin and Choi, Myungwoo and Jeon, Seokwoo},
    title = {{Graphene-based thermoelectric materials: toward sustainable energy-harvesting systems}},
    journal = {Chemical Communications},
    volume = {61},
    number = {27},
    pages = {5050-5063},
    year = {2025},
    month = {04},
    issn = {1359-7345},
    doi = {10.1039/d4cc06821a},
    url = {https://doi.org/10.1039/d4cc06821a},
}

@book{ziman, 
address={London}, 
title={{Electrons and Phonons: The Theory of Transport Phenomena}}, 
publisher={Oxford, Clanderon Press}, 
author={Ziman,  J. M.}, 
year={1960}
}

@article{HWANG2023467,
title = {Large scale graphene thermoelectric device with high power factor using gradient doping profile},
journal = {Carbon},
volume = {201},
pages = {467-472},
year = {2023},
issn = {0008-6223},
doi = {https://doi.org/10.1016/j.carbon.2022.09.048},
url = {https://www.sciencedirect.com/science/article/pii/S0008622322007722},
author = {Hyeon Jun Hwang and So-Young Kim and Sang Kyung Lee and Byoung Hun Lee}
}

@Article{ZT_1,
author="Hatef Sadeghi and Sara Sangtarash and Colin J. Lambert",
title="Enhancing the thermoelectric figure of merit in engineered graphene nanoribbons",
journal="Beilstein Journal of Nanotechnology",
year="2015",
volume="6",
pages="1176-1182",
issn="2190-4286",
doi="10.3762/bjnano.6.119",
copyright="Sadeghi et al; licensee Beilstein-Institut",
publisher="Beilstein-Institut",
URL="https://doi.org/10.3762/bjnano.6.119",
}

@article{ZT_2,
	author = {Sevin{\c c}li, H{\^a}ldun and Sevik, Cem and {\c C}a{\u g}ın, Tahir and Cuniberti, Gianaurelio},
	da = {2013/02/06},
	doi = {10.1038/srep01228},
	id = {Sevin{\c c}li2013},
	isbn = {2045-2322},
	journal = {Scientific Reports},
	number = {1},
	pages = {1228},
	title = {A bottom-up route to enhance thermoelectric figures of merit in graphene nanoribbons},
	ty = {JOUR},
	url = {https://doi.org/10.1038/srep01228},
	volume = {3},
	year = {2013}}

@Article{Dean2010,
author={Dean, C. R.
and Young, A. F.
and Meric, I.
and Lee, C.
and Wang, L.
and Sorgenfrei, S.
and Watanabe, K.
and Taniguchi, T.
and Kim, P.
and Shepard, K. L.
and Hone, J.},
title={Boron nitride substrates for high-quality graphene electronics},
journal={Nature Nanotechnology},
year={2010},
month={Oct},
day={01},
volume={5},
number={10},
pages={722-726},
issn={1748-3395},
doi={10.1038/nnano.2010.172},
url={https://doi.org/10.1038/nnano.2010.172}
}

@article{CBM_R2025,
  title = {{Thermoelectric transport in graphene under strain fields modeled by Dirac oscillators}},
  author = {Ca\~nas, Juan A. and Bonilla, Daniel A. and Mart\'{\i}n-Ruiz, A.},
  journal = {Phys. Rev. B},
  volume = {112},
  issue = {10},
  pages = {104206},
  numpages = {14},
  year = {2025},
  month = {Sep},
  publisher = {American Physical Society},
  doi = {10.1103/3r17-kfy7},
  url = {https://link.aps.org/doi/10.1103/3r17-kfy7}
}

@article{CBM_PB2026,
title = {Charge and energy transport in graphene with smooth finite-range disorder},
journal = {Physica B: Condensed Matter},
volume = {729},
pages = {418431},
year = {2026},
issn = {0921-4526},
doi = {https://doi.org/10.1016/j.physb.2026.418431},
url = {https://www.sciencedirect.com/science/article/pii/S0921452626001894},
author = {Juan A. Cañas and Daniel A. Bonilla and J.C. Pérez-Pedraza and A. Martín-Ruiz}
}

@article{DasSarmaRMP2011,
  title = {{Electronic transport in two-dimensional graphene}},
  author = {Das Sarma, S. and Adam, Shaffique and Hwang, E. H. and Rossi, Enrico},
  journal = {Rev. Mod. Phys.},
  volume = {83},
  issue = {2},
  pages = {407--470},
  numpages = {0},
  year = {2011},
  month = {May},
  publisher = {American Physical Society},
  doi = {10.1103/RevModPhys.83.407},
  url = {https://link.aps.org/doi/10.1103/RevModPhys.83.407}
}

@article{PeresRMP2010,
  title = {{Colloquium: The transport properties of graphene: An introduction}},
  author = {Peres, N. M. R.},
  journal = {Rev. Mod. Phys.},
  volume = {82},
  issue = {3},
  pages = {2673--2700},
  numpages = {0},
  year = {2010},
  month = {Sep},
  publisher = {American Physical Society},
  doi = {10.1103/RevModPhys.82.2673},
  url = {https://link.aps.org/doi/10.1103/RevModPhys.82.2673}
}

@article{NovoselovNature2005,
  title   = {Two-dimensional gas of massless Dirac fermions in graphene},
  author  = {Novoselov, K. S. and Geim, A. K. and Morozov, S. V. and Jiang, D. and Katsnelson, M. I. and Grigorieva, I. V. and Dubonos, S. V. and Firsov, A. A.},
  journal = {Nature},
  volume  = {438},
  number  = {7065},
  pages   = {197--200},
  year    = {2005},
  doi     = {10.1038/nature04233}
}

@article{CastroNetoRMP2009,
  title = {{The electronic properties of graphene}},
  author = {Castro Neto, A. H. and Guinea, F. and Peres, N. M. R. and Novoselov, K. S. and Geim, A. K.},
  journal = {Rev. Mod. Phys.},
  volume = {81},
  issue = {1},
  pages = {109--162},
  numpages = {0},
  year = {2009},
  month = {Jan},
  publisher = {American Physical Society},
  doi = {10.1103/RevModPhys.81.109},
  url = {https://link.aps.org/doi/10.1103/RevModPhys.81.109}
}

@article{ZuevPRL2009,
  title = {{Thermoelectric and Magnetothermoelectric Transport Measurements of Graphene}},
  author = {Zuev, Yuri M. and Chang, Willy and Kim, Philip},
  journal = {Phys. Rev. Lett.},
  volume = {102},
  issue = {9},
  pages = {096807},
  numpages = {4},
  year = {2009},
  month = {Mar},
  publisher = {American Physical Society},
  doi = {10.1103/PhysRevLett.102.096807},
  url = {https://link.aps.org/doi/10.1103/PhysRevLett.102.096807}
}

@article{WeiPRL2009,
  title = {{Anomalous Thermoelectric Transport of Dirac Particles in Graphene}},
  author = {Wei, Peng and Bao, Wenzhong and Pu, Yong and Lau, Chun Ning and Shi, Jing},
  journal = {Phys. Rev. Lett.},
  volume = {102},
  issue = {16},
  pages = {166808},
  numpages = {4},
  year = {2009},
  month = {Apr},
  publisher = {American Physical Society},
  doi = {10.1103/PhysRevLett.102.166808},
  url = {https://link.aps.org/doi/10.1103/PhysRevLett.102.166808}
}

@article{HwangPRL2007,
  title   = {Carrier Transport in Two-Dimensional Graphene Layers},
  author  = {Hwang, E. H. and Adam, S. and Das Sarma, S.},
  journal = {Physical Review Letters},
  volume  = {98},
  number  = {18},
  pages   = {186806},
  year    = {2007},
  doi     = {10.1103/PhysRevLett.98.186806}
}

@article{NovikovPRB2007,
  title = {{Elastic scattering theory and transport in graphene}},
  author = {Novikov, D. S.},
  journal = {Phys. Rev. B},
  volume = {76},
  issue = {24},
  pages = {245435},
  numpages = {17},
  year = {2007},
  month = {Dec},
  publisher = {American Physical Society},
  doi = {10.1103/PhysRevB.76.245435},
  url = {https://link.aps.org/doi/10.1103/PhysRevB.76.245435}
}

@article{WehlingPRL2010,
  title = {{Resonant Scattering by Realistic Impurities in Graphene}},
  author = {Wehling, T. O. and Yuan, S. and Lichtenstein, A. I. and Geim, A. K. and Katsnelson, M. I.},
  journal = {Phys. Rev. Lett.},
  volume = {105},
  issue = {5},
  pages = {056802},
  numpages = {4},
  year = {2010},
  month = {Jul},
  publisher = {American Physical Society},
  doi = {10.1103/PhysRevLett.105.056802},
  url = {https://link.aps.org/doi/10.1103/PhysRevLett.105.056802}
}

@article{FerreiraPRB2011,
  title = {{Unified description of the dc conductivity of monolayer and bilayer graphene at finite densities based on resonant scatterers}},
  author = {Ferreira, Aires and Viana-Gomes, J. and Nilsson, Johan and Mucciolo, E. R. and Peres, N. M. R. and Castro Neto, A. H.},
  journal = {Phys. Rev. B},
  volume = {83},
  issue = {16},
  pages = {165402},
  numpages = {22},
  year = {2011},
  month = {Apr},
  publisher = {American Physical Society},
  doi = {10.1103/PhysRevB.83.165402},
  url = {https://link.aps.org/doi/10.1103/PhysRevB.83.165402}
}

@article{WuPRB2014,
  title = {{Scattering of two-dimensional massless Dirac electrons by a circular potential barrier}},
  author = {Wu, Jhih-Sheng and Fogler, Michael M.},
  journal = {Phys. Rev. B},
  volume = {90},
  issue = {23},
  pages = {235402},
  numpages = {16},
  year = {2014},
  month = {Dec},
  publisher = {American Physical Society},
  doi = {10.1103/PhysRevB.90.235402},
  url = {https://link.aps.org/doi/10.1103/PhysRevB.90.235402}
}

@article{CaridadNatComm2016,
	author = {Caridad, Jos{\'e}M. and Connaughton, Stephen and Ott, Christian and Weber, Heiko B. and Krsti{\'c}, Vojislav},
	date = {2016/09/27},
	doi = {10.1038/ncomms12894},
	id = {Caridad2016},
	isbn = {2041-1723},
	journal = {Nature Communications},
	number = {1},
	pages = {12894},
	title = {{An electrical analogy to Mie scattering}},
	url = {https://doi.org/10.1038/ncomms12894},
	volume = {7},
	year = {2016}}

@article{KlosPRB2010,
  title = {{Effect of short- and long-range scattering on the conductivity of graphene: Boltzmann approach vs tight-binding calculations}},
  author = {K\l{}os, J. W. and Zozoulenko, I. V.},
  journal = {Phys. Rev. B},
  volume = {82},
  issue = {8},
  pages = {081414(R)},
  numpages = {4},
  year = {2010},
  month = {Aug},
  publisher = {American Physical Society},
  doi = {10.1103/PhysRevB.82.081414},
  url = {https://link.aps.org/doi/10.1103/PhysRevB.82.081414}
}

@article{ApplPhysA2024a,
	author = {Kakavandi, T. and Rezania, H.},
	date = {2024/04/23},
	doi = {10.1007/s00339-024-07512-9},
	id = {Kakavandi2024},
	isbn = {1432-0630},
	journal = {Applied Physics A},
	number = {5},
	pages = {337},
	title = {Impurity atoms effects on electronic properties and Seebeck coefficient of armchair graphene like nanoribbon},
	volume = {130},
	year = {2024}
}

@article{Crossno2016,
	author = {Jesse Crossno and Jing K. Shi and Ke Wang and Xiaomeng Liu and Achim Harzheim and Andrew Lucas and Subir Sachdev and Philip Kim and Takashi Taniguchi and Kenji Watanabe and Thomas A. Ohki and Kin Chung Fong},
	doi = {10.1126/science.aad0343},
	journal = {Science},
	number = {6277},
	pages = {1058-1061},
	title = {{Observation of the Dirac fluid and the breakdown of the Wiedemann-Franz law in graphene}},
	volume = {351},
	year = {2016}}

@article{Tu2023,
  title = {{Wiedemann-Franz law in graphene}},
  author = {Tu, Yi-Ting and Das Sarma, Sankar},
  journal = {Phys. Rev. B},
  volume = {107},
  issue = {8},
  pages = {085401},
  numpages = {13},
  year = {2023},
  month = {Feb},
  publisher = {American Physical Society},
  doi = {10.1103/PhysRevB.107.085401}
}

@article{PhysRevB.108.245415,
  title = {Wiedemann-Franz law in graphene in the presence of a weak magnetic field},
  author = {Tu, Yi-Ting and Das Sarma, Sankar},
  journal = {Phys. Rev. B},
  volume = {108},
  issue = {24},
  pages = {245415},
  numpages = {6},
  year = {2023},
  month = {Dec},
  publisher = {American Physical Society},
  doi = {10.1103/PhysRevB.108.245415},
  url = {https://link.aps.org/doi/10.1103/PhysRevB.108.245415}
}

@Article{ma14112704,
AUTHOR = {Rycerz, Adam},
TITLE = {Wiedemann–Franz Law for Massless Dirac Fermions with Implications for Graphene},
JOURNAL = {Materials},
VOLUME = {14},
YEAR = {2021},
NUMBER = {11},
ARTICLE-NUMBER = {2704},
URL = {https://www.mdpi.com/1996-1944/14/11/2704},
PubMedID = {34063902},
ISSN = {1996-1944},
DOI = {10.3390/ma14112704}
}

@article{Inglot2015,
  title = {Thermoelectric effect enhanced by resonant states in graphene},
  author = {Inglot, M. and Dyrda\l{}, A. and Dugaev, V. K. and Barna\ifmmode \acute{s}\else \'{s}\fi{}, J.},
  journal = {Phys. Rev. B},
  volume = {91},
  issue = {11},
  pages = {115410},
  numpages = {7},
  year = {2015},
  month = {Mar},
  publisher = {American Physical Society},
  doi = {10.1103/PhysRevB.91.115410},
  url = {https://link.aps.org/doi/10.1103/PhysRevB.91.115410}
}

@article{Feshbach1954,
  author = {Feshbach, H. and Porter, C. E. and Weisskopf, V. F.},
  title = {Model for nuclear reactions with neutrons},
  journal = {Phys. Rev.},
  volume = {96},
  pages = {448},
  year = {1954},
  doi = {10.1103/PhysRev.96.448}
}

@book{Mott,
  author = {Mott, N. F. and Massey, H. S. W.},
  title = {The Theory of Atomic Collisions},
  edition = {3rd},
  publisher = {Clarendon Press},
  address = {Oxford},
  year = {1965},
  chapter = {VIII},
  pages = {184}
}

@book{CantoHussein2013,
  author = {Canto, L. F. and Hussein, M. S.},
  title = {Scattering Theory of Molecules, Atoms and Nuclei},
  publisher = {World Scientific},
  address = {Singapore},
  year = {2013},
  chapter = {7},
  pages = {291}
}

@article{Jiang_APL2023,
    author = {Jiang, Wen-Cheng and Li, Jian and Li, Qing-Xu and Zhu, Jia-Ji},
    title = {The reciprocating and bipolar non-Hermitian skin effect engineered by spin–orbit coupling},
    journal = {Applied Physics Letters},
    volume = {123},
    number = {20},
    pages = {201107},
    year = {2023},
    month = {11},
    issn = {0003-6951},
    doi = {10.1063/5.0174400}
}

@article{Ghaemi_PRA2021,
  title = {Compatibility of transport effects in non-Hermitian nonreciprocal systems},
  author = {Ghaemi-Dizicheh, Hamed and Schomerus, Henning},
  journal = {Phys. Rev. A},
  volume = {104},
  issue = {2},
  pages = {023515},
  numpages = {11},
  year = {2021},
  month = {Aug},
  publisher = {American Physical Society},
  doi = {10.1103/PhysRevA.104.023515},
  url = {https://link.aps.org/doi/10.1103/PhysRevA.104.023515}
}

@article{kozii_Arx2017,
  title = {{Non-Hermitian topological theory of finite-lifetime quasiparticles: Prediction of bulk Fermi arc due to exceptional point}},
  author = {Kozii, Vladyslav and Fu, Liang},
  journal = {Phys. Rev. B},
  volume = {109},
  issue = {23},
  pages = {235139},
  numpages = {5},
  year = {2024},
  month = {Jun},
  publisher = {American Physical Society},
  doi = {10.1103/PhysRevB.109.235139},
  url = {https://link.aps.org/doi/10.1103/PhysRevB.109.235139}
}

@article{Shen_PRL2018,
  title = {Quantum Oscillation from In-Gap States and a Non-Hermitian Landau Level Problem},
  author = {Shen, Huitao and Fu, Liang},
  journal = {Phys. Rev. Lett.},
  volume = {121},
  issue = {2},
  pages = {026403},
  numpages = {6},
  year = {2018},
  month = {Jul},
  publisher = {American Physical Society},
  doi = {10.1103/PhysRevLett.121.026403},
  url = {https://link.aps.org/doi/10.1103/PhysRevLett.121.026403}
}

@article{Zyuzin_PRB2018,
  title = {Flat band in disorder-driven non-Hermitian Weyl semimetals},
  author = {Zyuzin, A. A. and Zyuzin, A. Yu.},
  journal = {Phys. Rev. B},
  volume = {97},
  issue = {4},
  pages = {041203(R)},
  numpages = {5},
  year = {2018},
  month = {Jan},
  publisher = {American Physical Society},
  doi = {10.1103/PhysRevB.97.041203},
  url = {https://link.aps.org/doi/10.1103/PhysRevB.97.041203}
}

@book{Moiseyev_B2011,
   title =     {Non-Hermitian Quantum Mechanics},
   author =    {Nimrod Moiseyev},
   publisher = {Cambridge University Press},
   isbn =      {0521889723; 9780521889728},
   year =      {2011},
   edition =   {1},
   url =       {libgen.li/file.php?md5=dac7693e865d5721bd16dde28ebb4c7c}}

@Book{Haug_B2008,
title="Quantum Kinetics in Transport and Optics of Semiconductors",
author="Haug, Hartmut and Jauho, Antti Pekka",
edition= "2",
year="2008",
publisher="Springer Berlin Heidelberg",
address="Berlin, Heidelberg",
pages="79--83",
isbn="978-3-540-73564-9"
}

@article{Rammer_RMP1986,
  title = {{Quantum field-theoretical methods in transport theory of metals}},
  author = {Rammer, J. and Smith, H.},
  journal = {Rev. Mod. Phys.},
  volume = {58},
  issue = {2},
  pages = {323--359},
  numpages = {0},
  year = {1986},
  month = {Apr},
  publisher = {American Physical Society},
  doi = {10.1103/RevModPhys.58.323},
  url = {https://link.aps.org/doi/10.1103/RevModPhys.58.323}
}

@article{Lee_APL2011,
    author = {Lee, Young Gon and Kang, Chang Goo and Jung, Uk Jin and Kim, Jin Ju and Hwang, Hyeon Jun and Chung, Hyun-Jong and Seo, Sunae and Choi, Rino and Lee, Byoung Hun},
    title = {Fast transient charging at the graphene/SiO2 interface causing hysteretic device characteristics},
    journal = {Applied Physics Letters},
    volume = {98},
    number = {18},
    pages = {183508},
    year = {2011},
    month = {05},
    issn = {0003-6951},
    doi = {10.1063/1.3588033}
}

@article{Wang_ACSN2010,
    author = {Wang, Haomin and Wu, Yihong and Cong, Chunxiao and Shang, Jingzhi and Yu, Ting},
    title = {Hysteresis of Electronic Transport in Graphene Transistors},
    journal = {ACS Nano},
    volume = {4},
    number = {12},
    pages = {7221-7228},
    year = {2010},
    doi = {10.1021/nn101950n},
    note ={PMID: 21047068}
}

@article{Li_PRL2012,
  title = {Femtosecond Population Inversion and Stimulated Emission of Dense Dirac Fermions in Graphene},
  author = {Li, T. and Luo, L. and Hupalo, M. and Zhang, J. and Tringides, M. C. and Schmalian, J. and Wang, J.},
  journal = {Phys. Rev. Lett.},
  volume = {108},
  issue = {16},
  pages = {167401},
  numpages = {5},
  year = {2012},
  month = {Apr},
  publisher = {American Physical Society},
  doi = {10.1103/PhysRevLett.108.167401}
}

@article{
Britnell_Sci2012,
author = {L. Britnell  and R. V. Gorbachev  and R. Jalil  and B. D. Belle  and F. Schedin  and A. Mishchenko  and T. Georgiou  and M. I. Katsnelson  and L. Eaves  and S. V. Morozov  and N. M. R. Peres  and J. Leist  and A. K. Geim  and K. S. Novoselov  and L. A. Ponomarenko },
title = {Field-Effect Tunneling Transistor Based on Vertical Graphene Heterostructures},
journal = {Science},
volume = {335},
number = {6071},
pages = {947-950},
year = {2012},
doi = {10.1126/science.1218461}}

@article{Auslender_PRB2007,
  title = {Generalized kinetic equations for charge carriers in graphene},
  author = {Auslender, M. and Katsnelson, M. I.},
  journal = {Phys. Rev. B},
  volume = {76},
  issue = {23},
  pages = {235425},
  numpages = {15},
  year = {2007},
  month = {Dec},
  publisher = {American Physical Society},
  doi = {10.1103/PhysRevB.76.235425}
}

@article{Peres_PRB2006,
  title = {Electronic properties of disordered two-dimensional carbon},
  author = {Peres, N. M. R. and Guinea, F. and Castro Neto, A. H.},
  journal = {Phys. Rev. B},
  volume = {73},
  issue = {12},
  pages = {125411},
  numpages = {23},
  year = {2006},
  month = {Mar},
  publisher = {American Physical Society},
  doi = {10.1103/PhysRevB.73.125411}
}

@article{Hwang_PRB2009,
  title = {Theory of thermopower in two-dimensional graphene},
  author = {Hwang, E. H. and Rossi, E. and Das Sarma, S.},
  journal = {Phys. Rev. B},
  volume = {80},
  issue = {23},
  pages = {235415},
  numpages = {5},
  year = {2009},
  month = {Dec},
  publisher = {American Physical Society},
  doi = {10.1103/PhysRevB.80.235415}
}

@article{Bonilla_PRB2026,
  title = {Non-Hermitian thermoelectric transport in graphene: Tunable anomalous transmission through complex barriers},
  author = {Bonilla, Daniel A. and Ca\~nas, Juan A. and P\'erez-Pedraza, J. C. and Mart\'{\i}n-Ruiz, A.},
  journal = {Phys. Rev. B},
  volume = {114},
  issue = {17},
  pages = {175407},
  numpages = {21},
  year = {2026},
  month = {Sep},
  publisher = {American Physical Society},
  doi = {10.1103/crjs-p5vv}
}

@article{Massicotte2021,
    author = {Massicotte, Mathieu and Soavi, Giancarlo and Principi, Alessandro and Tielrooij, Klaas-Jan},
    title = {{Hot carriers in graphene – fundamentals and applications}},
    journal = {Nanoscale},
    volume = {13},
    number = {18},
    pages = {8376-8411},
    year = {2021},
    month = {05},
    issn = {2040-3364},
    doi = {10.1039/d0nr09166a},
    url = {https://doi.org/10.1039/d0nr09166a}
}

@article{Francis2026,
    author = {Francis, Shinto
Mundackal and Mohonta, Sajib Kumar and Chiluwal, Shailendra and Sharma, Bipin and Rao, Rahul and Puneet, Pooja and Gong, Yu and Podila, Ramakrishna},
    title = {{Quantum Interference
and Localization in Disordered
Graphene}},
    journal = {ACS Nano},
    volume = {20},
    number = {7},
    pages = {5463-5475},
    year = {2026},
    month = {02},
    issn = {1936-0851},
    doi = {10.1021/acsnano.5c13512},
    url = {https://doi.org/10.1021/acsnano.5c13512}
}

@article{Lofwander2007,
  title = {{Impurity scattering and Mott's formula in graphene}},
  author = {L\"ofwander, Tomas and Fogelstr\"om, Mikael},
  journal = {Phys. Rev. B},
  volume = {76},
  issue = {19},
  pages = {193401},
  numpages = {4},
  year = {2007},
  month = {Nov},
  publisher = {American Physical Society},
  doi = {10.1103/PhysRevB.76.193401},
  url = {https://link.aps.org/doi/10.1103/PhysRevB.76.193401}
}

@article{Ashida2020,
	author = {Yuto Ashida and Zongping Gong and Masahito Ueda},
	doi = {10.1080/00018732.2021.1876991},
	journal = {Advances in Physics},
	number = {3},
	pages = {249--435},
	publisher = {Taylor \& Francis},
	title = {{Non-Hermitian physics}},
	volume = {69},
	year = {2020}}

@article{Muga2004,
	author = {J.G. Muga and J.P. Palao and B. Navarro and I.L. Egusquiza},
	doi = {https://doi.org/10.1016/j.physrep.2004.03.002},
	issn = {0370-1573},
	journal = {Physics Reports},
	number = {6},
	pages = {357-426},
	title = {{Complex absorbing potentials}},
	url = {https://www.sciencedirect.com/science/article/pii/S0370157304001218},
	volume = {395},
	year = {2004}}

@article{Sukhachov2020,
  title = {{Non-Hermitian impurities in Dirac systems}},
  author = {Sukhachov, P. O. and Balatsky, A. V.},
  journal = {Phys. Rev. Res.},
  volume = {2},
  issue = {1},
  pages = {013325},
  numpages = {13},
  year = {2020},
  month = {Mar},
  publisher = {American Physical Society},
  doi = {10.1103/PhysRevResearch.2.013325},
  url = {https://link.aps.org/doi/10.1103/PhysRevResearch.2.013325}
}

@article{Kokkinakis2026,
	author = {Kokkinakis, Emmanouil T. and Komis, Ioannis and Makris, Konstantinos G. and Economou, Eleftherios N.},
	date = {2026/03/14},
	doi = {10.1038/s42005-026-02558-y},
	id = {Kokkinakis2026},
	isbn = {2399-3650},
	journal = {Communications Physics},
	number = {1},
	pages = {152},
	title = {{Non-Hermitian impurity problem}},
	url = {https://doi.org/10.1038/s42005-026-02558-y},
	volume = {9},
	year = {2026}}

@article{Terh2023,
  title = {{Scattering dynamics and boundary states of a non-Hermitian Dirac equation}},
  author = {Terh, Yun Yong and Banerjee, Rimi and Xue, Haoran and Chong, Y. D.},
  journal = {Phys. Rev. B},
  volume = {108},
  issue = {4},
  pages = {045419},
  numpages = {11},
  year = {2023},
  month = {Jul},
  publisher = {American Physical Society},
  doi = {10.1103/PhysRevB.108.045419},
  url = {https://link.aps.org/doi/10.1103/PhysRevB.108.045419}
}

@article{Tzortzakakis2021,
  title = {{Transport and spectral features in non-Hermitian open systems}},
  author = {Tzortzakakis, A. F. and Makris, K. G. and Szameit, A. and Economou, E. N.},
  journal = {Phys. Rev. Res.},
  volume = {3},
  issue = {1},
  pages = {013208},
  numpages = {10},
  year = {2021},
  month = {Mar},
  publisher = {American Physical Society},
  doi = {10.1103/PhysRevResearch.3.013208},
  url = {https://link.aps.org/doi/10.1103/PhysRevResearch.3.013208}
}

@article{Li2025,
  title = {{Universal Non-Hermitian Transport in Disordered Systems}},
  author = {Li, Bo and Chen, Chuan and Wang, Zhong},
  journal = {Phys. Rev. Lett.},
  volume = {135},
  issue = {3},
  pages = {033802},
  numpages = {7},
  year = {2025},
  month = {Jul},
  publisher = {American Physical Society},
  doi = {10.1103/z9m1-3mwb},
  url = {https://link.aps.org/doi/10.1103/z9m1-3mwb}
}

@article{Brouwer1997,
  title = {{Voltage-probe and imaginary-potential models for dephasing in a chaotic quantum dot}},
  author = {Brouwer, P. W. and Beenakker, C. W. J.},
  journal = {Phys. Rev. B},
  volume = {55},
  issue = {7},
  pages = {4695--4702},
  numpages = {0},
  year = {1997},
  month = {Feb},
  publisher = {American Physical Society},
  doi = {10.1103/PhysRevB.55.4695},
  url = {https://link.aps.org/doi/10.1103/PhysRevB.55.4695}
}

@article{Visuri2022,
  title = {{Symmetry-Protected Transport through a Lattice with a Local Particle Loss}},
  author = {Visuri, A.-M. and Giamarchi, T. and Kollath, C.},
  journal = {Phys. Rev. Lett.},
  volume = {129},
  issue = {5},
  pages = {056802},
  numpages = {6},
  year = {2022},
  month = {Jul},
  publisher = {American Physical Society},
  doi = {10.1103/PhysRevLett.129.056802},
  url = {https://link.aps.org/doi/10.1103/PhysRevLett.129.056802}
}

@article{Uchino2022,
  title = {{Comparative study for two-terminal transport through a lossy one-dimensional quantum wire}},
  author = {Uchino, Shun},
  journal = {Phys. Rev. A},
  volume = {106},
  issue = {5},
  pages = {053320},
  numpages = {14},
  year = {2022},
  month = {Nov},
  publisher = {American Physical Society},
  doi = {10.1103/PhysRevA.106.053320},
  url = {https://link.aps.org/doi/10.1103/PhysRevA.106.053320}
}

@article{Visuri2023,
  title = {{Nonlinear transport in the presence of a local dissipation}},
  author = {Visuri, A.-M. and Giamarchi, T. and Kollath, C.},
  journal = {Phys. Rev. Res.},
  volume = {5},
  issue = {1},
  pages = {013195},
  numpages = {18},
  year = {2023},
  month = {Mar},
  publisher = {American Physical Society},
  doi = {10.1103/PhysRevResearch.5.013195},
  url = {https://link.aps.org/doi/10.1103/PhysRevResearch.5.013195}
}

@article{Gievers2024,
  title = {{Quantum wires with local particle loss: Transport manifestations of fluctuation-induced effects}},
  author = {Gievers, Marcel and M\"uller, Thomas and Fr\"oml, Heinrich and Diehl, Sebastian and Chiocchetta, Alessio},
  journal = {Phys. Rev. B},
  volume = {110},
  issue = {20},
  pages = {205419},
  numpages = {20},
  year = {2024},
  month = {Nov},
  publisher = {American Physical Society},
  doi = {10.1103/PhysRevB.110.205419},
  url = {https://link.aps.org/doi/10.1103/PhysRevB.110.205419}
}

@article{Sticlet2022,
  title = {{Kubo Formula for Non-Hermitian Systems and Tachyon Optical Conductivity}},
  author = {Sticlet, Doru and D\'ora, Bal\'azs and Moca, C\ifmmode \u{a}\else \u{a}\fi{}t\ifmmode \u{a}\else \u{a}\fi{}lin Pa\ifmmode \mbox{\c{s}}\else \c{s}\fi{}cu},
  journal = {Phys. Rev. Lett.},
  volume = {128},
  issue = {1},
  pages = {016802},
  numpages = {6},
  year = {2022},
  month = {Jan},
  publisher = {American Physical Society},
  doi = {10.1103/PhysRevLett.128.016802},
  url = {https://link.aps.org/doi/10.1103/PhysRevLett.128.016802}
}

@article{Yan2024,
  title = {{Transport theory in non-Hermitian systems}},
  author = {Yan, Qing and Li, Hailong and Sun, Qing-Feng and Xie, X. C.},
  journal = {Phys. Rev. B},
  volume = {110},
  issue = {4},
  pages = {045138},
  numpages = {13},
  year = {2024},
  month = {Jul},
  publisher = {American Physical Society},
  doi = {10.1103/PhysRevB.110.045138},
  url = {https://link.aps.org/doi/10.1103/PhysRevB.110.045138}
}

@article{Wei2025,
  title = {{Gauge invariant quantum transport theory for non-Hermitian systems}},
  author = {Wei, Miaomiao and Wang, Bin and Wang, Jian},
  journal = {Phys. Rev. B},
  volume = {111},
  issue = {7},
  pages = {075135},
  numpages = {9},
  year = {2025},
  month = {Feb},
  publisher = {American Physical Society},
  doi = {10.1103/PhysRevB.111.075135},
  url = {https://link.aps.org/doi/10.1103/PhysRevB.111.075135}
}

@article{Yang2026,
  title = {{Extended Landauer-B\"uttiker Formula for Current through Open Quantum Systems with Gain or Loss}},
  author = {Yang, Chao and Wang, Yucheng},
  journal = {Phys. Rev. Lett.},
  volume = {136},
  issue = {5},
  pages = {056304},
  numpages = {11},
  year = {2026},
  month = {Feb},
  publisher = {American Physical Society},
  doi = {10.1103/trty-58r2},
  url = {https://link.aps.org/doi/10.1103/trty-58r2}
}

@article{Kleger2026,
  title = {{Physical constraints on effective non-Hermitian systems}},
  author = {Kleger, Aaron and Boyack, Rufus},
  journal = {Phys. Rev. B},
  volume = {113},
  issue = {19},
  pages = {195136},
  numpages = {17},
  year = {2026},
  month = {May},
  publisher = {American Physical Society},
  doi = {10.1103/c4fn-g1m7},
  url = {https://link.aps.org/doi/10.1103/c4fn-g1m7}
}

@article{Niu2026,
	author = {Pengbin Niu and Li Xu and Hui Yao and Hong-Gang Luo},
	doi = {https://doi.org/10.1016/j.physe.2025.116385},
	issn = {1386-9477},
	journal = {Physica E: Low-dimensional Systems and Nanostructures},
	pages = {116385},
	title = {{Anomalous Seebeck effect in non-Hermitian double quantum dots}},
	url = {https://www.sciencedirect.com/science/article/pii/S1386947725002152},
	volume = {175},
	year = {2026}}
	
	
	
\end{document}